\documentclass[12pt]{article}
\usepackage[left=1in,right=1in,top=1.1in,bottom=1.1in]{geometry}
\usepackage{amsmath,amssymb}
\usepackage{supertabular}
\usepackage{booktabs}
\usepackage{graphicx}
\usepackage{floatrow}
\usepackage{latexsym}
\usepackage{array}
\usepackage{supertabular}
\usepackage{lastpage}
\usepackage{titlesec}
\usepackage{caption}
\usepackage{subcaption}
\usepackage{cite}

\makeatletter \@addtoreset{equation}{section} \makeatother

\newcommand{\pslash}{\partial\kern-.5 em/}
\newcommand{\Dslash}{D\kern-.7 em/}
\newcommand{\dslash}{d\kern-.5 em/}
\newcommand{\nslash}{\nabla\kern-.7 em/\kern.2em}
\newcommand{\tp}{\tilde\pslash}

\newcommand{\tD}{\tilde\Dslash}
\newcommand{\tn}{\tilde\nslash}
\newcommand{\tgamma}{\tilde\gamma}

\newcommand{\tphi}{\tilde\phi}
\newcommand{\Fslash}{F\kern-.7 em/}
\newcommand{\tF}{\tilde\Fslash}
\newcommand{\Aslash}{A\kern-.4em/}

\newcommand{\tlambda}{\tilde\lambda}
\newcommand{\tpsi}{\tilde\psi}
\newcommand{\kslash}{k\kern-.5 em/}
\newcommand{\bk}{\bar\kslash}
\newcommand\fft[2]{{\frac{#1}{#2}}}
\newcommand\ft[2]{{\textstyle\frac{#1}{#2}}}
\newcommand\nn{\nonumber}
\begin{document}

\begin{titlepage}

\begin{flushright}
MCTP-13-35
\end{flushright}

\vskip 0.5 cm

\begin{center}

{\Large \bf Small Treatise on Spin-3/2 Fields }\\

\vskip 0.5cm

{\Large \bf and their Dual Spectral Functions}
\vskip 1cm

{\large James T. Liu${}^1$, Leopoldo A. Pando Zayas${}^1$ and Zhenbin Yang${}^2$}

 \vskip.3cm

\end{center}

\vskip .4cm

\centerline{\it ${}^1$ Michigan Center for Theoretical
Physics}
\centerline{ \it Randall Laboratory of Physics, The University of
Michigan}
\centerline{\it Ann Arbor, MI 48109--1040}
\vskip .4cm

\centerline{\it ${}^2$ Department of Modern Physics}
\centerline{\it  University of Science and Technology of China}
\centerline{\it Hefei, China 230026}

\date{\today}

\begin{abstract}
In this work we systematically study various aspects of spin-3/2 fields in a curved background.  We mostly focus on a minimally coupled massive spin-3/2 field in arbitrary dimensions, and solve the equation of motion either explicitly or numerically in AdS, Schwarzschild-AdS and Reissner-Nordstr\"om-AdS backgrounds. Although not the main focus of this work, we also make a connection with the gravitino equation of motion in gauged supergravity. Motivated by the AdS/CFT correspondence, we emphasize calculational improvements and technical details of the dual spectral functions. We attempt to provide a coherent and comprehensive picture of the existing literature.
\end{abstract}
\end{titlepage}

%\maketitle
\vspace{-5mm}
\tableofcontents

\section{Introduction}

The Anti de-Sitter/Conformal Field Theory (AdS/CFT) correspondence \cite{Maldacena:1997re,Gubser:1998bc,Witten:1998qj,Aharony:1999ti} postulates an equivalence between a theory containing gravity on one side and a field theory on the other side.  This correspondence has motivated the study of classical fields in asymptotically Anti-de-Sitter (AdS) spacetimes in an unprecedented way. Typically, the correspondence is most powerful in the limit when gravity is weakly coupled, which corresponds to the strongly coupled regime in the dual field theory side. The duality has also recently found interesting applications to various aspects typical of condensed matter systems (see reviews  \cite{Hartnoll:2009sz,Herzog:2009xv,McGreevy:2009xe,Horowitz:2010gk,Sachdev:2011wg}). In this paper we aim at providing a systematic and largely comprehensive study of the spin-3/2 field in asymptotically AdS spacetimes  with a particular view towards its applications in the context of the Anti de-Sitter/Condensed Matter (AdS/CM) correspondence.

One of our motivations is the number of interesting and suggestive results that have been obtained recently by studying fermionic fields in asymptotically AdS spacetimes and reinterpreting the results from the dual strongly coupled field theory side. In particular, there have been beautiful results regarding the structure of the dual spectral function of fermions based on predictions from AdS/CFT \cite{Lee:2008xf,Faulkner:2009wj,Cubrovic:2009ye,Faulkner:2009am}, including an important approach to exciting condensed matter phenomena such as non-Fermi liquids  \cite{Lee:2008xf,Liu:2009dm} (see, e.g., the review \cite{Iqbal:2011ae} and references therein).

In the framework of the AdS/CFT correspondence, after initially finding interesting phenomena in some toy models of holography, it is paramount to consider the possibility of embedding into a ultraviolet complete theory like string theory. This approach has already proved fruitful in discussing aspects of holographic superconductors; in particular, it has provided an answer for the behavior in certain corners of the phase space \cite{Gauntlett:2009dn,Gauntlett:2009bh}.

Typically top-down models originating from string theory contain a spin-3/2 field which descends through dimensional reduction from the gravitino in 10 or 11 dimensions. The effective actions of the fermions in some UV-complete versions are known in some fairly general cases. For example, complete dimensional reductions of the fermionic sector were constructed from 10 and 11 dimensional supergravity on Sasaki-Einstein internal manifolds \cite{Bah:2010yt,Bah:2010cu,Liu:2011dw}. The resulting lower dimensional theory includes a coupled system of spin-3/2 and spin-1/2 fields. A complete analysis of such systems remains an interesting open problem. Some partial results in a  simplified context have already yielded powerful insights \cite{Edalati:2010ww,Edalati:2010ge} and point to the potential richness of new phenomena. Particular interest has also been paid to the spin-3/2 field in the context of supergravity theories where universal conclusions can be reached about the structure of the spectral functions \cite{Gauntlett:2011mf,Gauntlett:2011wm,DeWolfe:2011aa,DeWolfe:2012uv}.

One of the goals of this manuscript is to fill in and expand the number of results between toy models and supergravity models but to also consider general enough models as to help understand what aspects are intrinsic to spin-3/2 fields obtained from compactifications of supergravity models. Moreover, we aim at providing a unified picture by explicitly discussing putative aspects of gauge-fixing that make some approaches look very dissimilar when viewed separately; we also explicitly discuss the notational differences in various previous investigations. We explain how the difference between various calculations may be seen in group-theoretic terms.

The paper is organized as follows. In section \ref{EoM} we set up the problem of classically describing a spin-3/2 field. Section \ref{AdS} is devoted to reproducing results from the earlier literature and to the explicit calculation of the two-point correlator.  Section \ref{RN-AdS} discusses the case of a Reissner-Nordstr\"om-AdS (RN-AdS) background. We also consider the particular case of Schwarzschild-AdS (zero charge) black hole which is relevant for hydrodynamic questions and make contact with the existing literature.

Our conventions are summarized in Appendix~\ref{App:Conventions}.  In Appendix~\ref{App:flow}, we present a general formulation for obtaining the Green's function in terms of a first order flow equation based on a technique first introduced in \cite{Iqbal:2008by}. This analysis allows us to simplify the choice of boundary conditions, leading to better control in numerical evaluations.

%%%%%%%%%%%%%%%%%%%%%%%%%%%%%%%%%%%%%%%%%%%%%%%%%%%%%%%%%%%%%%%
\section{General equation for the Rarita-Schwinger field}\label{EoM}

We begin with a review of the Rarita-Schwinger spin-3/2 field \cite{Rarita:1941mf}, and reduce its equation of motion to a set of equations which we then proceed to solve in various backgrounds. This approach is effectively a decomposition of the spin-3/2 representation into spin-3/2 and spin-1/2 parts. We will return repeatedly to this decomposition from various points of view and will compare with similar choices made in the literature.

\subsection{The massless case}

Before proceeding to the more general case, it is worth reviewing the basic properties of the massless Rarita-Schwinger field.  The equation of motion for a massless spin-3/2 field propagating in a curved $D=d+1$ dimensional background is given by
\begin{equation}
\tilde\gamma^{MNP}\nabla_N\psi_P=0,
\label{eq:m=0eom}
\end{equation}
where $\nabla_N$ is the gravitational covariant derivative.  Acting on $\psi_P$, it takes the form
\begin{equation}
\nabla_N\psi_P=\partial_N\psi_P+\ft14\omega_N{}^{AB}\gamma_{AB}\psi_P-\Gamma^Q{}_{PN}\psi_Q,
\end{equation}
where $\omega_N$ is the spin connection and  $\Gamma^P{}_{LN}$ is the Levi-Civita connection.    It is often convenient to use the gamma matrix identity
\begin{equation}
{\tilde { \gamma } }^{M N P} = {\tilde { \gamma } }^{M}{\tilde { \gamma } }^{N}{\tilde { \gamma } }^{P}-{\tilde { \gamma } }^{M}{g}^{N P}-{\tilde { \gamma } }^{P}{g}^{M N}+{\tilde { \gamma } }^{N}{g}^{M P},
\label{eq:gmi}
\end{equation}
to rewrite (\ref{eq:m=0eom}) in the equivalent form
\begin{equation}
\tn\psi^M+(\tilde\gamma^M\tn-\nabla^M)\lambda-\tilde\gamma^M\beta=0,
\label{eq:m=0neweom}
\end{equation}
where we have defined
\begin{equation}
\lambda\equiv\tilde { \gamma  }^P \psi_P,\qquad  \beta\equiv \nabla^P \psi_P.
\label{eq:lambdabeta}
\end{equation}

Compared to the Dirac equation for spin-1/2, the spin-3/2 equation is complicated by having to handle the kinematics of a vector-spinor as well as gauge invariance (in the massless case).  In particular, since $\psi_M$ carries a vector as well as a spinor index of $\mathrm{SO}(1,d)$, it transforms in a reducible representation which is simply the sum of irreducible spin-1/2 and spin-3/2 components%
\footnote{Here we are a bit sloppy with the notation for $D$-dimensional spinor representations.  Heuristically, the decomposition is simply $\mathbf{\fft12}\times\mathbf1=\mathbf{\fft12}+\mathbf{\fft32}$ in terms of spin.}.
We now see that $\lambda$ is the spin-1/2 component of $\psi_M$, while the spin-3/2 component can be obtained by projection
\begin{equation}
\psi^{3/2}_M=\left(\delta_M^N-\fft1{d+1}\gamma_M\gamma^N\right)\psi_N=\psi_M-\fft1{d+1}\tilde\gamma_M\lambda.
\end{equation}
In addition, we may obtain a relation between $\lambda$ and $\beta$ by taking the gamma trace of (\ref{eq:m=0neweom})
\begin{equation}
(d-1)(\tn\lambda-\beta)=0\qquad\Rightarrow\qquad\beta=\tn\lambda
\label{eq:m=0gteom}
\end{equation}
(so long as $d>1$).

When the background is Ricci-flat, the massless equation (\ref{eq:m=0eom}) is invariant under the gauge transformation
\begin{equation}
\psi_P\to\psi_P+\nabla_P\zeta.
\label{eq:gaugexf}
\end{equation}
This may be seen by computing
\begin{equation}
\tilde\gamma^{MNP}\nabla_N\delta\psi_P=\tilde\gamma^{MNP}\nabla_N\nabla_P\zeta
=\ft12\tilde\gamma^{MNP}[\nabla_N,\nabla_P]\zeta=\ft12(R^{MN}-\ft12g^{MN}R)\tilde\gamma_N\zeta.
\end{equation}
This vanishes when the background solves the vacuum Einstein equations.  In this case, it is convenient to make the gauge choice $\lambda=0$.  The gamma traced equation of motion, (\ref{eq:m=0gteom}), then implies $\beta=0$, in which case the Rarita-Schwinger equation (\ref{eq:m=0neweom}) reduces to $\tn\psi^M=0$.  In Minkowski space, this is formally the same as imposing an independent Dirac equation for each vector component of the spin-3/2 field.  However, in a curved background the covariant derivative may act non-trivially on the vector index $M$ so that the equations would no longer be independent.

Note that the gauge choice $\lambda=0$ does not fix the gauge completely, as one may still make the shift (\ref{eq:gaugexf}) provided $\tn\zeta=0$.  It is instructive to see how this works in a flat background.  Working in momentum space, the gauge-fixed equation of motion becomes $\kslash\psi^M=0$.  Left-multiplying by $\kslash$ shows that $k^2=0$, which is simply the massless on-shell condition.  Specializing to $k^M=\omega(1,1,\vec0)$, we see that the Dirac-like equation is solved by taking the spinor polarization $\epsilon^M$ to satisfy the projection $P_+\epsilon^M=0$ where $P_\pm=\fft12(1\pm\gamma^{01})$.  The transverse ($\beta=0$) and gamma-traceless ($\lambda=0$) conditions then give $\psi^M=k^M\xi +\epsilon_t^M$, where $\xi$ is the longitudinal component satisfying $P_+\xi=0$ and $\epsilon_t^M$ is the transverse gamma-traceless component that satisfies $\gamma_M\epsilon^M=0$ and vanishes for $M=0,1$.  Finally, the residual gauge freedom may be used to set $\xi=0$.  Thus the physical on-shell degrees of freedom are given by
\begin{equation}
\psi^M=e^{i\omega(x^1-x^0)}\epsilon_t^M,\qquad\mbox{with}\qquad
P_+\epsilon_t^M=0,\quad \epsilon_t^0=0,\quad\epsilon_t^1=0,\quad\gamma_M\epsilon_t^M=0.
\end{equation}
In $D=4$, this gives the usual counting of two physical helicity states for a massless Majorana spin-3/2 particle.  Note that this discussion of gauge fixing and residual gauge symmetry is similar to that of the Maxwell field in Lorenz gauge.

\subsection{The minimally coupled massive case}

Having looked at the massless spin-3/2 equation, we now turn to the dynamics of a massive spin-3/2 field.  Allowing for minimal coupling to an abelian gauge field, the equation of motion takes the form%
\footnote{At this point we do not allow any non-minimal couplings to the gauge field. However, subsequently we will consider a Pauli coupling in the context of the gravitino in gauged supergravity.}
\begin{equation}
\label{Eq:RS}
{ \tilde { \gamma }  }^{M N L  } {D}_{N}{\psi}_{L}-{m}_{1}{g}^{M L}\psi_{L}-{m}_2{\tilde { \gamma } }^{M L}\psi_L=0,
\end{equation}
where
\begin{equation}
D_N=\nabla_N-ieA_N,
\end{equation}
is the gravitational and gauge covariant derivative.  Note that there are two possible covariant mass terms.  Just as in (\ref{eq:m=0neweom}), the spin-3/2 equation can be rewritten in the equivalent form
\begin{equation}
\label{eq:3/2eom}
(\tD-{ m }_{ 1 }+{ m }_{ 2 }){ \psi  }^{ M  }+({ \tilde { \gamma  }  }^{ M  }\tD-{ D  }^{ M  }-{ m }_{ 2 }{ \tilde { \gamma  }  }^{ M  })\lambda -{ \tilde { \gamma  }  }^{ M }\beta =0,
\end{equation}
where $\lambda$ takes the same form, but where $\beta$ is now defined as
\begin{equation}
\beta\equiv D^P\psi_P.
\end{equation}

Taking the gamma-trace of (\ref{eq:3/2eom}) allows us to obtain a relation between $\lambda$ and $\beta$
\begin{equation}
((d-1)\tD-{m}_1-d{m}_2)\lambda=(d-1)\beta.
\label{eq:gtrace}
\end{equation}
This generalizes (\ref{eq:m=0gteom}) of the massless case.  In addition to the gamma-trace, we can obtain a useful condition by acting with $D_M$ on (\ref{eq:3/2eom}).  After simplification, the result is
\begin{equation}
(m_2-m_1)\beta=\left(m_2\tD + \frac{H_{M N}\tilde{\gamma}^M \tilde{\gamma}^N}{2}\right)\lambda-H_{M N}\tilde{\gamma}^M \psi^N,
\label{eq:Dtrace}
\end{equation}
where
\begin{equation}
H_{M N}\equiv\frac{R_{M N}}{2}+ie F_{M N}
\end{equation}
arises from the commutator of two covariant derivatives.

For $m_1\neq 0$, we can use (\ref{eq:gtrace}) to eliminate $\beta$ from (\ref{eq:3/2eom}) and (\ref{eq:Dtrace}) to obtain:
\begin{eqnarray}
\label{eq:mc32eom}
(\tD-{ m }_{ 1 }+{ m }_{ 2 }){ \psi  }^{ M  }+\left(\frac{m_1+m_2}{d-1}{ \tilde { \gamma  }  }^{ M  }-{ D  }^{ M  }\right)\lambda&=&0,\\
\label{eq:mc12eom}
\left(\tD + \tilde m + \frac{H_{M N}\tilde \gamma^M \tilde\gamma^N}{2m_1}\right)\lambda - \frac{H_{M N}}{m_1}\tilde\gamma^M\psi^N &=&0,
\end{eqnarray}
where
\begin{equation}
\tilde m = \frac{(m_2-m_1)(m_2 d+m_1)}{(d-1)m_1}.
\end{equation}
For  $m_1=0$, the corresponding equation for $\lambda$ is replaced by:
\begin{equation}
\left(\frac{d m_2^2}{d-1}+\frac{H_{MN}\tgamma^M\tgamma^N}{2}\right)\lambda=H_{M N}\tgamma^M \psi^N.
\label{eq:m1=0eqn}
\end{equation}
In general, the strategy for solving this system is to first solve for $\lambda$, and then for $\psi^{N}$.

Consider, for example, the massive spin-3/2 equation in an Einstein background $R_{MN}=\Lambda g_{MN}$ with vanishing gauge field.  In this case, $H_{MN}=\fft12\Lambda g_{MN}$, and the equations of motion reduce to
\begin{eqnarray}
\label{eq:Rflatpsi}
(\tn-m_1+m_2)\psi^M&=&\left(\nabla^M-\fft{m_1+m_2}{d-1}\tilde\gamma^M\right)\lambda,\\
\left(\tn+\tilde m+\fft{d-1}{4m_1}\Lambda\right)\lambda&=&0.
\label{eq:Rflatlambda}
\end{eqnarray}
We can see that $\lambda$ propagates as a massive spin-1/2 degree of freedom, while also acting as a source to the $\psi^M$ equation of motion.  In order to extract the spin-3/2 component of $\psi^M$, we may perform the decomposition
\begin{equation}
\psi^M=\psi_t^M+\tilde\gamma^M\zeta+\nabla^M\xi,
\end{equation}
where $\psi_t^M$ is transverse gamma-traceless.  A formal expression for the spinors $\zeta$ and $\xi$ in terms of $\lambda$ and $\beta$ may be obtained by inverting the definitions (\ref{eq:lambdabeta})
\begin{eqnarray}
\zeta&=&\fft1{d\square+(d+1)\Lambda/4}(\square\lambda-\tn\beta),\nn\\
\xi&=&\fft1{d\square+(d+1)\Lambda/4}((d+1)\beta-\tn\lambda).
\end{eqnarray}
Using the relation (\ref{eq:gtrace}) along with the $\lambda$ equation of motion (\ref{eq:Rflatlambda}) allows us to write
\begin{eqnarray}
\zeta&=&\fft{m_1(m_1-m_2)}{d(m_1-m_2)^2+(d-1)^2\Lambda/4}\lambda,\nn\\
\xi&=&-\fft{(d-1)m_1}{d(m_1-m_2)^2+(d-1)^2\Lambda/4}\lambda,
\end{eqnarray}
so that
\begin{equation}
\psi^M=\psi_t^M+\fft{m_1}{d(m_1-m_2)^2+(d-1)^2\Lambda/4}\left((m_1-m_2)\tilde\gamma^M
-(d-1)\nabla^M\right)\lambda.
\end{equation}
Inserting this into (\ref{eq:Rflatpsi}) then gives the equation for the spin-3/2 degrees of freedom
\begin{equation}
(\tn-m_1+m_2)\psi_t^M=0.
\end{equation}

In the case that $m_1=0$, the $\lambda$ equation (\ref{eq:m1=0eqn}) becomes algebraic.  So long as $dm_2^2+(d-1)^2\Lambda/4\ne0$, we are constrained to take $\lambda=0$, in which case (\ref{eq:gtrace}) gives $\beta=0$ as well.  Thus the longitudinal component necessarily vanishes.  This is similar to what happens in the case of the massive Maxwell field.  If, on the other hand, $dm_2^2+(d-1)^2\Lambda/4=0$, then there is no constraint on $\lambda$.  This is the case of a `massless' spin-3/2 field in a cosmological background where gauge invariance is restored \cite{Deser:1977uq}%
\footnote{In fact, there is a rich structure of gauge invariance, degrees of freedom and unitarity bounds for massive spin-3/2 propagation in a cosmological background \cite{Deser:2000dz,Grassi:2000dm,Deser:2000de,Deser:2001us,Deser:2001dt,Duff:2002sm}.}.

This gives an explicit demonstration of how $\psi^M$ can be decomposed into a spin-1/2 component $\lambda$ and a spin-3/2 component $\psi_t^M$.  Furthermore, the masses of $\lambda$ and $\psi_t^M$ are in general different, and are given by $|\tilde m+(d-1)\Lambda/4m_1|$ and $|m_2-m_1|$, respectively (in the case $m_1\ne0$). Having turned the spin-3/2 equation into a set of Dirac-like equations facilitates the intuitive understanding of the behavior of the fields.
One conceptual goal of this work is to clarify the group-theoretic origin of this decomposition and its implications for the dual field theory in terms of dimensions of operators.

%%%%%%%%%%%%%%%%%%%%%%%%%%%%%%%%%%%%%%%%%%%%%%%%%%%%%%%%%%%%%%%%%%%%%
\subsection{Supergravity and the gravitino equation of motion}

As we have seen above, the spin-3/2 field has a gauge symmetry in the case $m_1=0$ and $m_2=\fft12(d-1)\sqrt{-\Lambda/d}$.  It is in fact this particular case that corresponds to the massless gravitino in supergravity, and the gauge symmetry is nothing but local supersymmetry
\cite{Freedman:1976xh,Deser:1976eh,Townsend:1977qa,Deser:1977uq}.  Since many applications of AdS/CFT can be investigated in the supergravity limit of a full ultraviolet complete theory, it is worth expanding on the gravitino equation as both a particular case and extension by Pauli terms of the general spin-3/2 field equation (\ref{Eq:RS}).

Since the supergravity multiplet depends on dimension as well as the superalgebra, it is not possible to present a completely generic description of gravitino dynamics.  However, we can focus on a truncated subsector of many supergravity theories where only a background metric and gauge field are turned on.  In particular, what we have in mind is a Dirac gravitino charged under an abelian graviphoton.  The truncated gauged supergravity Lagrangian is then given by
\begin{equation}
e^{-1}\mathcal L=R+d(d-1)m^2-\ft14F_{MN}^2+\bar\psi_M\tilde\gamma^{MNP}\mathcal D_N\psi_P
+\cdots,
\end{equation}
where the supercovariant derivative is
\begin{equation}
\mathcal D_N=D_N+ia(\tilde\gamma_N{}^{AB}-2(d-2)\delta_N^A\tilde\gamma^B)F_{AB}
+\fft{m}2\tilde\gamma_N+\cdots.
\label{eq:scd}
\end{equation}
In the above, the ellipses denote terms that are present and are needed for closure of the algebra but that we are not concerned about.  The constants $a$ and $m$ are fixed by supersymmetry and are given by
\begin{equation}
a =\frac{1}{4\sqrt{2(d-1)(d-2)}}, \qquad
m=\fft4{\sqrt{2(d-1)(d-2)}}e,
\label{Eq:Couplings}
\end{equation}
where $e$ is the gauge coupling constant.  Note that $m$ is the inverse radius of the vacuum AdS space, and $\Lambda=-dm^2$.

The equation of motion for the gravitino is simply
\begin{equation}
\label{Eq:Gravitino}
\tgamma^{M N P}{\cal D}_N \psi_P=0.
\end{equation}
This is invariant under supersymmetry transformations $\psi_P\to\psi_P+\delta\psi_P$ where
\begin{equation}
\delta\psi_P=\mathcal D_P\varepsilon,
\end{equation}
so long as the integrability condition
\begin{equation}
\tgamma^M[{\mathcal D}_M,{\mathcal D_N}] \varepsilon=0,
\end{equation}
is satisfied.  This constrains the background $(g_{MN},A_M)$ to satisfy the Einstein-Maxwell equations of motion
\begin{eqnarray}
&&R_{MN}=\fft12\left(F_{MP}F_N{}^P-\fft1{2(d-1)}g_{MN}F^2\right)-dm^2g_{MN},\nn\\
&&dF=0,\qquad d*F=0.
\end{eqnarray}
There is actually a subtlety that arises here since we are potentially working with an incomplete set of fields.  In $D=4$, these bosonic equations of motion are complete, while in $D=5$, the Maxwell equation must be replaced by $d*F=\fft1{\sqrt3}F\wedge F$.  In both of these cases, the field content $(g_{MN},A_M,\psi_M)$ is complete for $\mathcal N=2$ gauged supergravity.  As this is no longer true for $D>5$, in higher dimensions we must impose the constraint $F_{M[N}F_{PQ]}=0$ for consistency in the truncated background.

The equation of motion (\ref{Eq:Gravitino}) can be rewritten using the gamma matrix identity (\ref{eq:gmi}).  After commuting $\mathcal D_N$ past gamma matrices when appropriate, we may arrive at
\begin{eqnarray}
0&=&(\tgamma^N\mathcal D_N-2ia\tF-m)\psi^M-4ia(\tgamma^M\tgamma^A-(d-1)g^{MA})
F_{AB}\psi^B\nn\\
&&+(\tgamma^M\tgamma^N\mathcal D_N-\mathcal D^M+2ia(d-1)(\tgamma^M\tF
-2F^M{}_B\tgamma^B)-(d-1)m\tgamma^M)\lambda-\tgamma^M\hat\beta,\qquad
\label{eq:gtoeom}
\end{eqnarray}
where
\begin{equation}
\hat\beta\equiv\mathcal D_M\psi^M.
\end{equation}
Taking the gamma trace of this equation gives after some manipulation
\begin{equation}
0=(d-1)[(\tgamma^M\mathcal D_M+2ia(d-2)\tF-dm)\lambda-\hat\beta].
\label{eq:gtogtrace}
\end{equation}
To solve the system, we may now make use of the gauge invariance of the gravitino to work in $\lambda=0$ gauge.  In this case, (\ref{eq:gtogtrace}) gives the condition $\hat\beta=0$, and we are left with the first line of (\ref{eq:gtoeom}) to solve.  As in the case of a massless spin-3/2 field in a Ricci-flat background, the $\lambda=0$ gauge fixing is incomplete, and we are left with the freedom to perform residual gauge transformations $\psi_P\to\psi_P+\mathcal D_P\zeta$, so long as $\zeta$ solves the equation $\tilde\gamma^M\mathcal D_M\zeta=0$.

In order to make connection with previous investigations of the gravitino \cite{Gauntlett:2011mf,Gauntlett:2011wm}, we may left-multiply the supercovariant derivative (\ref{eq:scd}) by $\tgamma^N$.  The result is
\begin{equation}
\tgamma^N\mathcal D_N=\tD-ia(d-3)\tF+\fft{m}2(d+1).
\end{equation}
Similarly, we find the relation
\begin{equation}
\hat\beta=\beta+2ia(d-3)F_{AB}\tgamma^A\psi^B+(ia\tF+\ft12m)\lambda.
\end{equation}
This allows us to write the gauge-fixed equations of motion as
\begin{eqnarray}
0&=&\left(\tD-ia(d-1)\tF+\fft{m}2(d-1)\right)\psi^M+4ia(\tgamma^A\tgamma^M+(d-3)g^{MA})
F_{AB}\psi^B,\nn\\
0&=&\beta+2ia(d-3)F_{AB}\tgamma^A\psi^B,
\end{eqnarray}
along with the gauge fixing constraint $\lambda=0$.  This agrees with the $D=4$ gravitino equation in \cite{Gauntlett:2011mf,Gauntlett:2011wm}, so long as the graviphoton is scaled up by a factor of two.

For later convenience, we introduce
\begin{equation}
m_2=\frac{m}2(d-1), \qquad g=a(d-1).
\label{eq:m2gdef}
\end{equation}
In this case, the gravitino equation can be written as
\begin{equation}
\left(\tD-ig\tF+m_2\right)\psi^M+4ia(\tgamma^A\tgamma^M+(d-3)g^{MA})
F_{AB}\psi^B=0.
\end{equation}
Note that $m_2$ corresponds directly to the second mass term in (\ref{Eq:RS}), while $g$ is the coefficient of the diagonal Pauli term.  For vanishing background gauge field, this reduces to $(\tn+m_2)\psi^M=0$, with $\lambda=0$ and $\beta=0$, which agrees with (\ref{eq:Rflatpsi}) found above (with $m_1=0$).

%%%%%%%%%%%%%%%%%%%%%%%%%%%%%%%%%%%%%%%%%%%%%%%%%%%%%%%%%%%%%%%%%%
\section{Solving the Rarita-Schwinger equation}\label{AdS}
%%%%%%%%%%%%%%%%%%%%%%%%%%%%%%%%%%%%%%%%%%%%%%%%%%%%%

Although we are ultimately interested in solving the equations of motion in a charged background, we first review the dynamics of the Rarita-Schwinger field in a pure AdS geometry.  After this, we will set up the more general system.  Solutions and applications to the holographic spectral function will be considered in subsequent sections.

\subsection{The solution in a pure AdS background}

The equations of motion in an Einstein background are given above in (\ref{eq:Rflatpsi}) and (\ref{eq:Rflatlambda}).  After defining
\begin{equation}
m_\pm=m_1\pm m_2,\qquad M=\fft{d(d-1)}{4m_1L^2}-\tilde m,
\label{eq:mmmpMdef}
\end{equation}
where $L$ is the AdS$_{d+1}$ radius, they may be rewritten as
\begin{equation}
(\tn-m_-)\psi^M=\left(\nabla^M-\fft{m_+}{d-1}\tgamma^M\right)\lambda,\qquad
(\tn-M)\lambda=0.
\label{eq:AdSeom}
\end{equation}
This is essentially the decomposition of the massive Rarita-Schwinger field into spin-1/2 and spin-3/2 degrees of freedom.  However, when $m_1=0$, the $\lambda$ equation is replaced by the gamma-traceless condition $\lambda=0$.%
\footnote{Although $\lambda$ becomes unconstrained when $m_2=\pm(d-1)/(2L)$, this is exactly where gauge invariance is restored.  In this case, $\lambda=0$ may be thought of as a choice of gauge.}
In this case, the spin-1/2 degree of freedom is either absent or can be gauged away.

Solutions to the equations of motion in an AdS background can be classified under representations of the AdS group $\mathrm{SO}(2,d)$.  For spin-3/2 and spin-1/2 given by (\ref{eq:AdSeom}), we have
\begin{equation}
E_0(\psi^M_t)=\fft{d}2+|m_-L|,\qquad E_0(\lambda)=\fft{d}2+|ML|.
\end{equation}
Note that the supergravity gravitino has $m_-=\pm(d-1)/2L$.  In this case, $E_0(\psi_t^M)=d-1/2$, while $\lambda$ becomes unphysical.

To be explicit, we work in the Poincar\'e patch of AdS$_{d+1}$ and take a metric of the form
\begin{equation}
ds^2=\frac{L^2}{r^2}(-dt^2+d\vec{x}_{d-1}^{\,2}+dr^2).
\end{equation}
The $\lambda$ equation then takes the form
\begin{equation}
\left(\tp-\fft{d}{2r}\tgamma^r-M\right)\lambda=0,
\end{equation}
where the $\tgamma^r$ term arises from the spin connection.  It is convenient to remove this term by rescaling $\lambda=(r/L)^{d/2}\hat\lambda$, so that
\begin{equation}
(\tp-M)\hat\lambda=0.
\label{eq:AdSde}
\end{equation}
As shown in Appendix~\ref{App:Conventions}, this rescaling generalizes to the black hole case.  The spin-3/2 equation splits according to $M=\{\mu,r\}$
\begin{eqnarray}
\left(\tp-\frac{1}{r}\tgamma^r-m_-\right)\hat\psi^\mu&=&\left(g^{\mu\nu}\partial_\nu
-\fft1{2r}\tgamma^\mu\tgamma^r-\frac{m_+}{d-1}\tgamma^\mu\right)\hat\lambda
+ \frac{1}{r}\tgamma^\mu\hat\psi^r,\nn\\
\left(\tp-\frac{2}{r}\tgamma^r-m_-\right)\hat\psi^r&=&g^{rr}\left(\partial_r+\frac{d-2}{2 r}
-\frac{m_+}{d-1} \tgamma_r\right)\hat\lambda,
\label{eq:AdSde2}
\end{eqnarray}
where we have used the same rescaling on $\psi^M$, namely $\psi^M=(r/L)^{d/2}\hat\psi^M$.

Working in $d$-dimensional momentum space allows (\ref{eq:AdSde}) and (\ref{eq:AdSde2}) to be written as a coupled set of first order equations
\begin{eqnarray}
\left(\gamma^r\partial_r+i\kslash-\fft{ML}r\right)\hat\lambda&=&0,\nn\\
\left(\gamma^r\left(\partial_r-\fft1r\right)+i\kslash-\fft{m_-L}r\right)\hat\psi^{\bar r}&=&
\left(\partial_r+\fft1r\left(\fft{d-2}2-\fft{m_+L}{d-1}\gamma^r\right)\right)\hat\lambda,\nn\\
\left(\gamma^r\partial_r+i\kslash-\fft{m_-L}r\right)\hat\psi^{\bar\mu}&=&
\left(ik^\mu-\fft1r\gamma^{\mu}\left(\fft12\gamma^r+\fft{m_+L}{d-1}\right)\right)\hat\lambda
+\fft1r\gamma^\mu\hat\psi^{\bar r}.
\label{Eq:AdS-3-2}
\end{eqnarray}
Note that we have transformed the vector index on the gravitino into tangent space, $\hat\psi^{\bar M}=(L/r)\hat\psi^M$.  The strategy for solving these equations is to solve them in order, first for $\hat\lambda$, then for $\hat\psi^{\bar r}$ and finally for $\hat\psi^{\bar\mu}$, using the previous solutions as sources.  While $\hat\lambda$ appears to be independent, we must in fact impose the constraint $\hat\lambda=\gamma_r\hat\psi^{\bar r}+\gamma_\mu\hat\psi^{\bar\mu}$ on the solution.

%%%%%%%%%%%%%%%%%%%%%%%%%%%%%%%%%%%%%%%%%%%%%%%%%%%%%%%%%%%%%%%%%%%%%%%%%%%%%%%%%%
\subsubsection{Connection to previous work}
%%%%%%%%%%%%%%%%%%%%%%%%%%%%%%%%%%%%%%%%%%%%%%%%%%%%%%%%%%%%%%%%%%%%%%%%%%%%%%%%%%%
The question of spin-3/2 fields was considered in the context of the AdS/CFT correspondence right after its inception. Some of the pioneering works in this direction include Refs.~\cite{Corley:1998qg,Volovich:1998tj} which first considered the problem, Refs.~\cite{Koshelev:1998tu,Matlock:1999fy} which allow for a massive Rarita-Schwinger field and Ref.~\cite{Rashkov:1999ji} which discussed the role of boundary terms. Many of these works elaborated on the spin-1/2 case developed in Ref~\cite{Mueck:1998iz}. In the context of the AdS/CFT correspondence the natural focus is on the computation of the 2-point correlator.

The equations of motion in an AdS$_{d+1}$ background, (\ref{eq:AdSde2}), reproduce those presented originally in Ref.~\cite{Matlock:1999fy}, provided we transform from $\hat\psi^{\bar M}$ back to $\psi^{\bar M}$ and take the AdS radius $L$ to be unity.  When $m_1=0$, $\hat\lambda$ becomes non-dynamical, and ought to be set to zero (either as a consequence of the equations of motion, or as a gauge choice).  In this case, the equations (\ref{Eq:AdS-3-2}) reduce to
\begin{eqnarray}
\left(\gamma^r\left(\partial_r-\fft1r\right)+i\kslash+\fft{m_2L}r\right)\hat\psi^{\bar r}&=&0,\nn\\
\left(\gamma^r\partial_r+i\kslash+\fft{m_2L}r\right)\hat\psi^{\bar\mu}&=&
\fft1r\gamma^\mu\hat\psi^{\bar r}.
\end{eqnarray}
Note that these may be combined as
\begin{equation}
\left(\gamma^r\partial_r+i\kslash+\fft{m_2L}r\right)\hat\psi^{\bar M}=
\fft1r\gamma^M\hat\psi^{\bar r}.
\end{equation}
This matches the form of the equation that was given in Ref.~\cite{Volovich:1998tj}.  The supergravity gravitino equation in an AdS background is obtained by taking $m_2L=\pm(d-1)/2$.

\subsubsection{The explicit solution}

As mentioned above, the strategy for solving the system of equations (\ref{Eq:AdS-3-2}) is to solve them sequentially.  We start with the homogeneous equations, which can all be brought into the form
\begin{equation}
\left(\gamma^r\partial_r+i\kslash-\fft\nu{r}\right)\zeta=0.
\end{equation}
By projecting with
\begin{equation}
P_\pm=\fft12(1\pm\gamma^r),
\label{Eq:Projectors}
\end{equation}
we may split this into two coupled first order equations
\begin{equation}
\left(\pm\partial_r-\fft\nu{r}\right)\zeta_\pm=-i\kslash\zeta_\mp.
\end{equation}
The solution is given by
\begin{equation}
\zeta=\sqrt{z}\left[\left(J_{\nu-\fft12}(z)+i\hat\kslash J_{\nu+\fft12}(z)\right)
\zeta_+^{(1)}
+\left(Y_{\nu-\fft12}(z)+i\hat\kslash Y_{\nu+\fft12}(z)\right)\zeta_+^{(2)}\right],
\end{equation}
where
\begin{equation}
z=\kappa r\quad\mbox{and}\quad\hat k^\mu=\fft{k^\mu}\kappa,\qquad\mbox{where}\qquad
\kappa=\sqrt{-k_\mu k^\mu}.
\end{equation}
Here $\zeta_+^{(1)}$ and $\zeta_+^{(2)}$ are two independent constant spinors with positive $\gamma^r$ eigenvalues.  Although we have chosen to parameterize the solution in terms of positive $\gamma^r$ spinors, we could equally well have chosen the opposite by taking $\zeta_+^{(i)}=i\hat\kslash\zeta_-^{(i)}$.

The particular solutions with the inhomogeneous terms on the right hand sides of (\ref{Eq:AdS-3-2}) are somewhat more difficult to obtain, but were worked out in Ref.~\cite{Matlock:1999fy}.  The solution may be given in terms of three constant spinor parameters, $a_+$, $b_+^\mu$ and $c_+$, and takes the form \cite{Matlock:1999fy}
\begin{eqnarray}
\tlambda&=&z^{\frac{d+1}{2}}\left(J_{ML-1/2}(z)+i\hat\kslash J_{ML+1/2}(z)\right)a_+,\\
\psi^{\bar r}&=&z^{\frac{d+3}{2}}\left(J_{m_-L-1/2}(z)+i\hat\kslash J_{m_-L+1/2}(z)\right)c_+\nn\\
&&+z^{\frac{d+1}{2}}\left(-(\mu_1+\mu_3 i\hat\kslash z)J_{ML-1/2}(z)
+(\mu_2 i\hat\kslash+\mu_3z)J_{ML+1/2}(z)\right)a_+,\\
\psi^{\bar\mu}&=&z^{\frac{d+1}{2}}\left(J_{m_-L-1/2}(z)+i\hat\kslash J_{m_-L+1/2}(z)\right)
b_+^\mu\nn\\
&&+z^{\frac{d+1}{2}}\left(-i\hat k^\mu i\hat\kslash zJ_{m_-L-1/2}(z)
+\left(-\gamma^\mu+i\hat k^\mu\left((2m_-L+1)\hat\kslash+z\right)\right)J_{m_-L+1/2}(z)\right)
c_+\nn\\
&&+z^{\frac{d+1}{2}}\Biggl(\left(\frac{\mu_1+\frac{m_+L}{d-1}+\frac{1}{2}}{(M+m_-)L}\gamma^\mu
-\mu_3i\hat{k}^\mu z\right)J_{ML-1/2}(z)\nn\\
&&\kern4em-\left(\frac{\mu_2-\frac{m_+L}{d-1}+\fft12}{(M+m_-)L}\gamma^\mu
+\mu_3i\hat{k}^\mu z\right)i\hat\kslash J_{ML+1/2}(z)\Biggr)a_+,
\label{eq:psimusoln}
\end{eqnarray}
where
\begin{equation}
\mu_1=\frac{\frac{m_+L}{d-1}-ML-\frac{d}{2}+1}{(M-m_-)L-1}, \qquad
\mu_2=\frac{\frac{m_+L}{d-1}-ML+\frac{d}{2}-1}{(M-m_-)L+1}, \qquad
\mu_3=\frac{1+\mu_1+\mu_2}{(M+m_-)L}.
\end{equation}
To avoid overly long expressions, we have written out the solution in terms of the Bessel function $J_\nu(z)$.  However, $J_\nu$ can be replaced by any linear combination of $J_\nu$ and $Y_\nu$, with the appropriate combination fixed by boundary conditions.  In general, for positive $M$ and $m_-$, $J_\nu$ corresponds to the normalizable solution, while $Y_\nu$ corresponds to the non-normalizable one.

Of course, we must still impose the gamma-trace condition $\lambda=\gamma_\mu\psi^{\bar \mu}+\gamma_r\psi^{\bar r}$.  This yields the restrictions
\begin{equation}
c_+=\frac{2}{d-1-2m_-L}i\hat k \cdot b_+, \qquad
\gamma\cdot b^+=0,
\end{equation}
so in particular $c_+$ is not an independent parameter.  Note that consistency of this gamma-trace condition is not automatic, but requires use of the relation (\ref{eq:mmmpMdef}) that ties together the spin-1/2 and spin-3/2 parts of the Rarita-Schwinger field.  In particular, one needs the  identities \cite{Matlock:1999fy}
\begin{equation}
\frac{\mu_1+\frac{m_+L}{d-1}+\frac{1}{2}}{(M+m_-)L}=\fft{1+\mu_1}d,\qquad
\frac{\mu_2-\frac{m_+L}{d-1}+\fft12}{(M+m_-)L}=-\fft{1+\mu_2}d.
\end{equation}
In any case, the result is that the physical spin-1/2 modes are parameterized by $a_+$ and the spin-3/2 modes by ($d$-dimensional) gamma-traceless $b_+^\mu$.  When $m_1=0$, or as a gauge choice for the massless gravitino, we must set $a_+=0$, in which case only spin-3/2 propagates.

Finally, note that, while the term proportional to $c_+$ in (\ref{eq:psimusoln}) is not particularly symmetrical in $J_{m_-L\mp1/2}$, this is merely an artifact of the parametrization.  The term can be symmetrized by performing the shift
\begin{equation}
b_+^\mu\to b_+^\mu-\left(\fft12\gamma^\mu i\hat\kslash+m_-Li\hat k^\mu\right)c_+.
\end{equation}
However, this shifted $b_+^\mu$ is no longer gamma-traceless, which makes it somewhat more awkward to work with.

%%%%%%%%%%%%%%%%%%%%%%%%%%%%%%%%%%%%%%%%%%%%%%%%%%%%%%%%%%%%%%%%%%%%%%%%%%%%%%%%%%%%%%%%%%%%%%
\subsection{The solution in more general backgrounds}
%%%%%%%%%%%%%%%%%%%%%%%%%%%%%%%%%%%%%%%%%%%%%%%%%%%%%%%%%%%%%%%%%%%%%%%%%%%%%%%%%%%%%%%%%%%%%%%

We are, of course, interested in the Rarita-Schwinger field in more general backgrounds, although we restrict to asymptotically AdS space times in the Poincar\'e patch.  Assuming isotropy of the transverse directions, we take a background metric of the form
\begin{equation}
\label{Eq:Metric}
ds^2=-e^{2A}dt^2+e^{2B}d\vec{x}_{d-1}^{\,2}+e^{2C}dr^2,
\end{equation}
where $A(r)$, $B(r)$ and $C(r)$ are functions that depend only on the radial coordinate $r$.  Although one of these functions can be removed by reparametrization of the $r$, we find it convenient to keep all three functions, so we can make direct contact to solutions in the literature.  In addition, we allow for a background radial electric field
\begin{equation}
A=A_t(r) dt\qquad\Rightarrow\qquad F=-A'_t\, dt\wedge dr,
\end{equation}
which is compatible with the isometries of the metric.  (Primes denote derivatives with respect to $r$.)  More details about aspects of this metric can be found in Appendix~\ref{App:Conventions}.

Since this background in general breaks $d$-dimensional Lorentz invariance, it is natural to separate out the time and radial components of $\psi^M$.  In particular, we take $M=\{t,i,r\}$.  Then, in this background, the minimally coupled equations of motion, (\ref{eq:mc32eom}) and (\ref{eq:mc12eom}) take the form
\begin{equation}
(\tp-ieA_t\tgamma^t + B' \tilde \gamma^r -m_-)\hat\psi^i=\left(g^{ij}(\partial_j
+\ft12B'\tgamma_j\tgamma^r) -\frac{m_+}{d-1} \tilde\gamma^i\right)\hat\lambda
- B' \tilde\gamma ^i \hat\psi^r,
\label{eq:spin3/2gen}
\end{equation}
and
\begin{eqnarray}
&&\kern-2em(\tp-ieA_t\tgamma^t + A' \tilde \gamma^r  - m_-)\hat\psi^t =
\left(g^{tt}(\partial_t-ieA_t+\ft12A'\tgamma_t\tgamma^r)-\frac{m_+}{d-1} \tilde\gamma^t \right)
\hat\lambda - A' \tilde\gamma ^t \hat\psi^r,\nn\\
&&\kern-2em(\tp-ieA_t\tgamma^t+ (B'+C')\tilde\gamma^r -m_-)\hat\psi^r =
\left(g^{rr}(\partial_r-\ft12(A'+(d-3)B'))-\frac{m_+}{d-1}\tilde\gamma^r\right)\hat\lambda\nn\\
&&\kern28em +g^{rr}(A'-B')\tilde\gamma_t\hat\psi^t,\nn\\
&&\kern-2em\left(\tp-ieA_t\tgamma^t+ \tilde m + \fft1{2m_1}(\ft12R-R_x^x+2ieA_t'\tgamma^r\tgamma^t)
\right)\hat\lambda\nn\\
&&\kern4em=\fft1{2m_1}\left(((R_t^t-R_x^x)\tgamma_t+2ieA_t'\tilde\gamma^r)\hat\psi^t
+((R_r^r-R_x^x)\tgamma_r-2ieA_t'\tgamma^t)\hat\psi^r\right),
\label{eq:spin1/2gen}
\end{eqnarray}
where the Ricci components $R_t^t$, $R_x^x$ and $R_r^r$ and the Ricci scalar $R$ are given in Appendix~\ref{App:Conventions}.  As above, and as detailed in this appendix, we have rescaled to new  fields $\hat\lambda$ and  $\hat\psi$ in order to remove the spin connection from the covariant derivative.

From a $(d-1)$-dimensional point of view, the components of $\psi^M$ are decomposed as three spin-1/2 fields $\psi^t$, $\psi^r$ and $\lambda$, and one spin-3/2 field $\psi^i$.  The equations of motion (\ref{eq:spin1/2gen}) mix the spin-1/2 components, and will in principle have to be solved simultaneously.  The equation for $\psi^i$ is given in (\ref{eq:spin3/2gen}), and involves $\hat\lambda$ and $\hat\psi^r$ as sources.  So long as we are only interested in spin-3/2 as far as the transverse components are concerned, we may set $\hat\lambda=\hat\psi^t=\hat\psi^r=0$.  We are then left with a homogeneous Dirac equation for $\hat\psi^i$:
\begin{equation}
(\tp-ieA_t\tgamma^t -m_-)\hat\psi^{\bar i}=0,
\end{equation}
where we have transformed into tangent space.  In $d$-dimensional momentum space, this becomes:
\begin{equation}
(\tilde\gamma^r \partial_r+i\tilde \gamma^i k_i -i\tilde \gamma^t (\omega+e A_t )-m_-)\hat\psi^{\bar i}=0,
\end{equation}
along with the constraint
\begin{equation}
\gamma^i \hat\psi_{\bar i}=0,
\end{equation}
which follows from demanding $\hat\lambda=0$.

Following \cite{Liu:2009dm}, we split $\psi^{\bar i}$ into definite eigenstates under the $\gamma^r$ projection (\ref{Eq:Projectors}).  The resulting equation takes the form
\begin{equation}
e^{B-C}(\partial_r\mp e^C m)\hat\psi^{\bar i}_{\pm}=\mp i \gamma^\mu k_\mu \hat\psi^{\bar i}_{\mp},
\label{eq:minc3/2}
\end{equation}
where $m=m_1-m_2$ and
\begin{equation}
k_\mu\equiv(-u,k_i) \quad \mbox{with} \quad  u=e^{B-A}(\omega+e A_t).
\label{eq:minc3/2k}
\end{equation}
Since the individual $\bar i$ components of $\hat\psi^{\bar i}$ satisfy an ordinary Dirac equation, the analysis of \cite{Liu:2009dm} carries over directly to the present case.

\subsubsection{Gravitino field}

The gravitino equation of motion differs from the minimally coupled one discussed above by the inclusion of a Pauli term.  In addition, we may use gauge invariance of the gravitino to work in $\lambda=0$ gauge.  In this case, the system decomposes into a coupled set of equations for $\hat\psi^t$ and $\hat\psi^r$ along with an equation for $\hat\psi^i$ with $\hat\psi^t$ and $\hat\psi^r$ as sources.

Focusing, as above, on the transverse spin-3/2 components, we set $\hat\psi^t=\hat\psi^r=0$
(along with $\lambda=0$ as a gauge choice).  The gravitino equation then reduces to
\begin{equation}
(\tp-ieA_t\tgamma^t+m_2 -ig\tF)\hat\psi^{\bar i}=0,
\end{equation}
where $m_2$ and $g$ are given in (\ref{eq:m2gdef}).  Using the projectors $P_\pm $ introduced in (\ref{Eq:Projectors}), we finally arrive at:
\begin{equation}
e^{B-C}(\partial_r\mp e^C m)\hat\psi^{\bar i}_{\pm}=\mp i \gamma^\mu k^\mp_\mu \hat\psi^{\bar i}_{\mp},
\label{eq:gtino3/2}
\end{equation}
where $m=-m_2$ and
\begin{equation}
k^\pm_\mu=(-u^\pm,k_i)\quad\mbox{where}\quad u^\pm=e^{B-A}(\omega+e A_t\pm 2 g F_{tr} e^{-C}).
\label{eq:gtino3/2k}
\end{equation}
As we are working in $\lambda=0$ gauge, we must also enforce the condition
\begin{equation}
\gamma^i \hat\psi_{\bar i}=0,
\end{equation}
on the solution.  Since this reduces to the minimally coupled case, (\ref{eq:minc3/2}) and (\ref{eq:minc3/2k}) in the limit $g\to0$, we will take this as the general form of the spin-3/2 equation in the subsequent analysis.

At this point, it is worth contrasting our approach of focusing on the transverse spin-3/2 components of the gravitino with that of Refs.~\cite{Policastro:2008cx,Gauntlett:2011mf,Gauntlett:2011wm}.  Given the interest of Ref.~\cite{Policastro:2008cx} on the field theory supercharge which relates to the time component of the gravitino, there $\psi^t$ was not set to zero, and instead the transverse spin-3/2 components were made to vanish.  In particular, Ref.~\cite{Policastro:2008cx} worked with $D=5$ and decomposed the five-dimensional gravitino into helicity components $\pm1/2$ and $\pm3/2$.  For propagation in the $z$ direction, there are four components in the helicity $\pm1/2$ sector: $\psi_t$, $\psi_r$, $\psi_z$ and $\psi=\gamma^x\psi_x+\gamma^y\psi_y$, and one in the helicity $\pm3/2$ sector: $\eta=\gamma^x\psi_x-\gamma^y\psi_y$.  The helicity $\pm3/2$ states are removed by setting $\eta=0$.  Then, by using the equations of motion and making the $\lambda=0$ gauge choice, it is possible to focus on the $\psi_t$ and $\psi_z$ sector.  This is essentially the opposite of what we have done above.
The analysis of Refs.~\cite{Gauntlett:2011mf,Gauntlett:2011wm} is similar.  However, they work in $D=4$, and focus on the coupled $\psi_t$ and $\psi_r$ sector.  The remaining components $\psi_x$ and $\psi_y$ can then be recovered by working out the consequences of the transverse and gamma-traceless conditions on $\psi_M$.

%%%%%%%%%%%%%%%%%%%%%%%%%%%%%%%%%%%%%%%%%%%%%%%%%%%%%%%%%%%%%%%%%%%%%%%%%%%%%%%%
\section{Spin-3/2 fields in a Reissner-Nordstr\"om background and holographic spectral functions}\label{RN-AdS}
%%%%%%%%%%%%%%%%%%%%%%%%%%%%%%%%%%%%%%%%%%%%%%%%%%%%%%%%%%%%%%%%%%%%%%%%%%%%%%%%%%

We now turn to the case of a charged black hole in asymptotically AdS spacetimes (RN-AdS). First, we analyze the  general situation and later on we consider the Schwarzschild solution as a limit.
The general metric of the planar RN-AdS black hole can be taken to be
\begin{equation}
ds^2=\frac{L^2}{r^2}\left(-f dt^2+d\vec x^2+\frac{dr^2}{f}\right), \qquad  f=1-(1+Q^2)\left(\frac{r}{r_0}\right)^d+ Q^2\left(\frac{r}{r_0}\right)^{2d-2},
\end{equation}
and the gauge field is:
\begin{equation}
A_t=\mu\left(1-\left(\frac{r}{r_0}\right)^{d-2}\right), \qquad Q=\sqrt{\frac{d-2}{2(d-1)}}\frac{\mu r_0}{L}.
\end{equation}
We can use a rescaling $x^M\rightarrow r_0 x^M$ to eliminate $r_0$.   (Later on we just set $r_0=1$.)
Then the result will be:\par
\begin{eqnarray}
&&f=1-(1+Q^2)r^d+ Q^2r^{2d-2}, \qquad
Q=\sqrt{\frac{d-2}{2(d-1)}}\frac{\mu r_0}{L},\nn\\
&&A_t=\mu(1-r^{d-2}), \qquad
F_{tr}=(d-2)\mu r^{d-3},
\end{eqnarray}
with
\begin{equation}
e^A=\frac{L}{r}\sqrt f, \qquad  e^B=\frac{L}{r}, \qquad  e^C=\frac{L}{r \sqrt f},
\end{equation}
in the notation of (\ref{Eq:Metric}). The zero temperature limit corresponds to $Q^2=d/(d-2)$.

%%%%%%%%%%%%%%%%%%%%%%%%%%%%%%%%%%%%%%%%%%%%%%%%%%%%%%%%%%%%%%%%%%%%%%%%%%%%%%
%\noindent {\Large\textbf {Boundary Conditions: Horizon and Asymptotics}}\par

In order to compute the holographic spin-3/2 Green's function. we must pay attention to the boundary conditions at the horizon and at the AdS boundary.  In particular, the horizon conditions determine the type of Green's function that is computed \cite{Iqbal:2009fd}, and the regarded Green's function is given by taking in-falling boundary conditions at the horizon.

We first consider the horizon, located at $r=1$.  In order to highlight the horizon behavior, we introduce $z=1-r$ and expand for small $z$.  For finite temperature and non-vanishing frequency, $\omega \neq 0$, we have
\begin{eqnarray}
&&f\rightarrow \nu z, \qquad  e^A\rightarrow L\sqrt{\nu z}, \qquad  e^B\rightarrow L, \qquad e^C\rightarrow \frac{L}{\sqrt {\nu z}},\nn\\
&&A_t\rightarrow (d-2)\mu z, \qquad F_{tr}\rightarrow (d-2) \mu,\nn\\
&&u^\pm \rightarrow \frac{\omega}{\sqrt{\nu z}}, \qquad e^{B-C} \rightarrow \sqrt {\nu z}, \qquad \kslash^\pm\rightarrow -\gamma^0 \frac{\omega}{\sqrt {\nu z}},
\end{eqnarray}
where $\nu = d-Q^2(d-2)>0$.  Note, in particular, that $\omega\ne0$ yields the dominant behavior at the horizon.  In the case, the general equation (\ref{eq:gtino3/2}) becomes
\begin{equation}
\sqrt{\nu z}\left(-\partial_z \mp \frac{L m}{\sqrt{\nu z}}\right)\Psi_\pm=\pm i \gamma^0 \frac{\omega}{\sqrt {\nu z}} \Psi_\mp,
\end{equation}
where $\Psi$ denotes any single component of $\hat\psi^{\bar i}$.  If we let $x=\sqrt z$ and $\omega_0=2 \omega/\nu$, we obtain
\begin{equation}
\left(x\partial_x\pm\frac{2 m L x}{\sqrt \nu}\right)\Psi_\pm=\mp i \gamma^0 \omega_0 \Psi_\mp.
\end{equation}
The solutions are:\par
\begin{eqnarray}
\Psi_+=A x^{i \omega_0}+ B x^{-i \omega_0},\nn\\
\Psi_-=C x^{i \omega_0}+ D x^{-i \omega_0},
\end{eqnarray}
where
\begin{equation}
A=-\gamma^0 C, \quad B=\gamma^0 D.
\end{equation}

An analysis of the boundary conditions at the horizon for real-time computations involving spinors was presented in \cite{Iqbal:2009fd} and we follow that work. Imposing in-falling boundary condition at the horizon means that we choose $B$ and $D$ to be non-vanishing. For zero temperature and $\omega \neq 0$, the boundary condition is still $B=\gamma^0 D$.

We will also be interested in the case of zero temperature and vanishing frequency, $\omega = 0$. This limit is particularly interesting from the point of view of applications to zero temperature condensed matter as it resembles the situation around quantum critical points. Recall that quantum critical points are generically defined as continuous phase changes of matter at absolute zero. Indeed, this intuition was confirmed in  \cite{Liu:2009dm} which found critical quantum behavior of Dirac field.

For zero temperature and $\omega=0$, the frequency no longer dominates at the horizon, and the spatial momentum becomes important.  We may exploit rotational symmetry to put the momentum in the $x^1$ direction.  We then find
\begin{eqnarray}
&&f\rightarrow d(d-1) z^2,\quad e^A\rightarrow L\sqrt{d(d-1)}z,\quad e^B\rightarrow L , \quad e^C\rightarrow \frac{L}{\sqrt {d(d-1} )z},\nn\\
&&A_t\rightarrow(d-2)\mu z, \quad F_{tr}\rightarrow (d-2) \mu,\nn\\
&&u^\pm \rightarrow e L \sqrt 2 \pm 2 g \sqrt{2 d (d-1)}, \quad e^{B-C} \rightarrow \sqrt {d(d-1)}z, \quad \kslash^\pm\rightarrow -\gamma^0 u^\pm+ \gamma^1 k_1,\qquad
\label{eq:T=0w=0lim}
\end{eqnarray}
where
\begin{equation}
\mu = \frac{\sqrt {2 d (d-1)}}{d-2}L
\end{equation}
is fixed by the zero temperature condition.  In this case, the spin-3/2 equation (\ref{eq:gtino3/2}) becomes
\begin{equation}
(\sqrt{d(d-1)}z \partial_z \pm mL )\Psi_\pm=\pm i \kslash^\mp \Psi_\mp.
\end{equation}
The solution takes the form:
\begin{equation}
\Psi_+=\begin{pmatrix}
    y_+^+x^{\frac{\kappa_1}{\sqrt{d(d-1)}}}+y_+^-x^{-\frac{\kappa_1}{\sqrt{d(d-1)}}} \\
    z_+^+x^{\frac{\kappa_2}{\sqrt{d(d-1)}}}+z_+^-x^{-\frac{\kappa_2}{\sqrt{d(d-1)}}} \\
\end{pmatrix},\qquad
\Psi_-=\begin{pmatrix}
    y_-^+x^{\frac{\kappa_2}{\sqrt{d(d-1)}}}+y_-^-x^{-\frac{\kappa_2}{\sqrt{d(d-1)}}} \\
    z_-^+x^{\frac{\kappa_1}{\sqrt{d(d-1)}}}+z_-^-x^{-\frac{\kappa_1}{\sqrt{d(d-1)}}} \\
\end{pmatrix},
\end{equation}
where
\begin{equation}
\label{Eq:Frequencies}
\kappa_1=\sqrt{(k-u^-)(k+u^+)+(m L)^2}, \quad  \kappa_2=\sqrt{(k+u^-)(k-u^+)+(m L)^2},
\end{equation}
and where $y_\pm^\pm,z_\pm^\pm$ are constants.
To impose in-falling  boundary conditions at the horizon, we have to demand that either $\kappa_1$ or $\kappa_2$ be imaginary.  After choosing the in-going wave condition, we end up with a relation between $y_\pm$ and $z_\pm$:
\begin{eqnarray}
\nonumber i(k+u^+) y_+&=&(\sqrt{(k-u^-)(k+u^+)+(m L)^2}+m L) z_-, \\
i (k-u^+) z_+&=&(\sqrt{(k+u^-)(k-u^+)+(m L)^2}+m L) y_-.
\end{eqnarray}
At finite temperature the functions $u^\pm$ in (\ref{eq:T=0w=0lim}) simply reduce to $u^\pm = \pm{2 g (d-2) \mu}/{L}$.

We now turn to the asymptotic behavior of the solution near the AdS boundary.  Since the boundary corresponds to $r\to0$, we expand for small $r$:
\begin{eqnarray}
&&f \rightarrow 1, \qquad e^A\sim e^B\sim e^C\rightarrow\frac{L}{r}, \qquad A_t \rightarrow \mu,\nn\\
&&F_{tr}e^{-C}\rightarrow (d-2)\mu r^{d-2}/L, \qquad  u^\pm \rightarrow \bar\omega=\omega + e \mu,  \qquad \kslash^\pm \rightarrow\bk.\\
\end{eqnarray}
The equation of motion (\ref{eq:gtino3/2}) takes the following form:
\begin{equation}
\left(\partial_r \mp \frac{L m}{r}\right)\Psi_\pm=\mp i \bk \Psi_\mp,
\end{equation}
whose solution is: \par
\begin{eqnarray}
\Psi_+=D r^{mL}+ C r^{1-mL}, \qquad  \Psi_-=A r^{-mL} + B r^{mL+1},
\end{eqnarray}
with \par
\begin{equation}
C=\frac{i \bk}{2mL-1}A, \qquad  B=\frac{i \bk}{2mL+1}D, \qquad  \bar k_\mu=(-(\omega+ e \mu),k_i).
\end{equation}

The field equations can be simplified further by exploiting the symmetries of the problem. For example, we can always perform a Lorentz transform to limit our consideration to only $\omega$ and the $k_1$ component.   (See Appendix~\ref{App:Conventions}  for our choice of gamma matrices.) Introducing
\begin{equation}
\Psi_\pm=\left(
                      \begin{array}{c}
                        y_\pm \\
                        z_\pm \\
                      \end{array}
                    \right), \,\,\,{\rm and}\,\,\,\,
 \xi_+=\frac{i y_+}{z_-}, \quad  \xi_-=-\frac{i z_+}{y_-},
\end{equation}
changes the original equation into
\begin{eqnarray}
\nonumber e^{B-C}(\partial_r\mp e^C m)y_\pm=\mp i (k-u^\mp)z_\mp, \\
e^{B-C}(\partial_r\mp e^C m)z_\pm=\mp i (k+u^\mp)y_\mp,
\end{eqnarray}
along with a flow equation for $\xi_\pm$:
\begin{equation}
\label{Eq:FlowRN}
e^{B-C}\partial_r \xi_\pm=\pm (k_1\mp u^-)+ 2 e^B m \xi_\pm \mp (k_1\pm u^+)\xi_\pm^2.
\end{equation}
The horizon condition in the case of finite temperature or $\omega \neq 0$ becomes:
\begin{eqnarray}
\xi_\pm\mid_{r\rightarrow 1}=i.
\end{eqnarray}
For the case of $T=0$ and $\omega=0$, this is replaced by
\begin{equation}
\xi_\pm\mid_{r\rightarrow 1}=\frac{m L + \sqrt{(k\mp u^-)(k\pm u^+)+ m^2 L ^2}}{u^+\pm k}.
\label{eq:omega=0hc}
\end{equation}
The retarded Green's function is, therefore,  defined as the limiting value for an initial value problem:
\begin{eqnarray}
G_R=\lim _{ \epsilon \rightarrow 0 }\left.{\epsilon^{-2mL}\left(
             \begin{array}{cc}
               \xi_+ & 0  \\
               0 & \xi_-   \\
             \end{array}
           \right)} \right|_{r=\epsilon},
\end{eqnarray}
where one should extract the finite terms in the limit.

Finally, note that there is a symmetry:
\begin{equation}
G_{22}(\omega,k)=G_{11}(\omega,-k).
\end{equation}
For the case with no gauge field and $m=0$, there is an extra relation:
\begin{equation}
G_{22}(\omega,k)=-\frac{1}{G_{11}(\omega,k)}.
\end{equation}
These symmetries were pointed out in \cite{Liu:2009dm} for spin-1/2 fields; here we simply remark that such symmetries extend to the spin-3/2 case.

%%%%%%%%%%%%%%%%%%%%%%%%%%%%%%%%%%%%%%%%%%%%%%
\subsection{Spectral functions in Black Hole-AdS}

With the general analysis out of the way, we are finally able to focus on the spectral function of the operators dual to spin-3/2 fields. In general, the spectral function of an operator is a measure of the density of states which couple to the operator. It is proportional to the imaginary part of the retarded Green's function of the operator in momentum space,  $A(\omega,\vec k)=-2\mbox{\,Im\,} G_R(\omega, \vec k)$.

The flow equation (\ref{Eq:FlowRN}) is suitable for numerically computing  Green's functions. Here we list some example results, where we fix $d=3$, $L=1$, $\mu=1$ and $e=1/2$ unless otherwise noted.  For the uncharged background (i.e.\ Schwarzschild-AdS), we do not need $\mu$ nor $e$, and for the exploration of quantum critical behavior, we set $\mu=2\sqrt{3}$ as required by the zero temperature limit).

%%%%%%%%%%%%%%%%%%%%%%%%%%%%%%%%%%%%%%%%%%%%%%
\subsubsection{Spectral functions in Schwarzschild-AdS}

\begin{figure}[htp]
\includegraphics[width=0.5\textwidth]{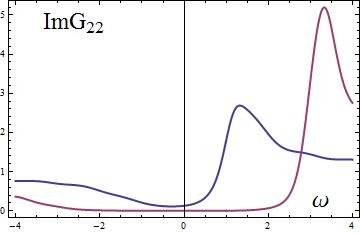}\\
\caption{The imaginary part of the Green's function for Schwarzschild-AdS with $m=0$.  The blue line corresponds to $k=1$, and the red one to $k=3$.}\label{Fig:Sch_RS}
\end{figure}

We first consider Schwarzschild-AdS with $m=0$.  (Recall that for the general R-S field  $m=m_1-m_2$, and for the gravitino $m=-m_2$.)  A typical graph of the spectral function is given in Figure~\ref{Fig:Sch_RS}, where we have taken the spatial momenta to be $k=1$ and $k=3$.  There are some salient features of this graph that we would like to highlight. First, the spectral function is nonnegative and dictated by unitarity. Second, there is a peak in the graph that indicates the presence of a quasi-particle in this setup. Third, as $\omega \to \infty$, the spectral function goes to 1.

\begin{figure}
\includegraphics[width=0.5\textwidth]{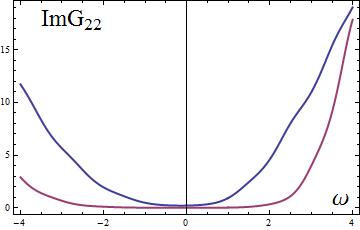}\\
\caption{Gravitino spectral function in Schwarzschild AdS ($m=1$). The blue line corresponds to $k=1$, and the red one to $k=3$.}\label{Fig:Sch_Gra}
\end{figure}

We may contrast the $m=0$ case shown in Figure~\ref{Fig:Sch_RS} with the $m=1$ case for the gravitino shown in Figure~\ref{Fig:Sch_Gra}. (The value $m=1$ is demanded by supersymmetry and gauge invariance of the gravitino.)  The main new feature in Figure~\ref{Fig:Sch_Gra} is that the peak is no longer visible. In fact, the peak is hidden, or overwhelmed by the fact that we have a nonvanishing mass, $m\neq 0$. More importantly, in all situations with $m\neq 0$ we find a divergent behavior in the spectral functions; this is easily seen in the asymptotic behavior. This behavior is rather typical from the conformal field theory point of view. Namely,  it  follows as a direct consequence of the form of the two-point function in a conformal field theory.
\begin{eqnarray}
\mbox{\,Im\,} G_R\propto \omega^{2\nu},
\end{eqnarray}
where $\nu$ is related to the effective mass of the gravity field or, equivalently, to the dimension of the corresponding operator (see Appendix \ref{App:flow} for an explicit derivation).

Downplaying the distinction between spin-1/2 and spin-3/2 fields, the peak in Figure~\ref{Fig:Sch_RS} is generally known as corresponding to the phonino pole in the Green's function. As argued in Refs.~\cite{Lebedev:1989rz,Lebedev:1988sq}, a supersymmetric thermal medium must have a fermionic collective excitation, that was denoted the phonino. Subsequent studies focused  mostly on supersymmetric models and the late-time tails \cite{Lebedev:1988sq}. An important contribution setting the holographic calculation for fermionic quasiparticles was put forward in  \cite{Kovtun:2003vj}. One of our main conclusions, which we emphasize throughout, consists in the similarities between spin-3/2 and the already well studied spin-1/2 case.

The definitive susy hydrodynamics (gravitino) treatment in Schwarzschild-AdS in its modern characterization within  AdS/CFT was presented in  \cite{Policastro:2008cx}. This work holographically computed the dispersion relation for a hydrodynamic mode of fluctuation (the phonino) of the density of supersymmetry current in ${\cal N} = 4$ SYM at strong coupling. The mode appears as a pole at
low frequency and momentum in the correlator of supercurrents; the speed and coefficient of attenuation were also presented.  More recently, the diffusion constant and the supercharge density has been calculated in  \cite{Kontoudi:2012mu}. Interesting recent work extending the previous analysis to the supergravity backgrounds corresponding to M2 and M5 branes and considering also other  transport (supercharge diffusion constants in supersymmetric field theories) has been developed \cite{Erdmenger:2013thg}.

\subsubsection{Spectral functions in RN-AdS}

Let us briefly recall one of the main motivations for the study of spectral functions of field theory operators via the AdS/CFT correspondence. One of the main entries in the AdS/CFT dictionary states that considering the boundary theory at finite temperature
and finite density corresponds to putting a black
hole in the bulk geometry. We further consider a quantum field theory which contains
fermions charged under a global $U(1)$ symmetry and allow for a finite $U(1)$ charge density to be introduced into such a theory. In this manuscript we focus on the spectral functions for fermionic operators dual to the spin-3/2 field in the gravity side.

\begin{figure}[htp]
\includegraphics[width=0.5\textwidth]{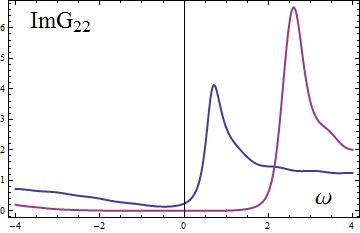}\\
\caption{Spectral function in RN-AdS with $m=0$, and $g=0$. The blue line corresponds to $k=1$, and the red one to $k=3$.}\label{Fig:RN_Gra}
\end{figure}

For the RN-AdS background, we  restore the couplings and the gauge field to calculate the Green's Function. For the general spin-3/2 field we do not include the Pauli coupling; namely, we set $g=0$. A typical representative of spectral function in this case is given in Figure~\ref{Fig:RN_Gra}, where we have set $m=0$ and considered two values of the spatial momenta, $k=1$ and $3$.
As in the case of the spectral function for Schwarzschild-AdS, here, for RN-AdS, we find some generic properties of the spectral function.
The spectral function is nonnegative as implied by unitarity. There is also a peak, signaling a pole which is identifiable with a quasi-particle, we also note that the value of the peak grows linearly with  $k$. Finally, for $\omega \to \infty$, the spectral function goes to 1.

Motivated by the interesting results of \cite{Edalati:2010ww,Edalati:2010ge} but with the overall goal of capturing aspects of the full problem of the fermionic sector in reductions of supergravity theories \cite{Bah:2010yt,Bah:2010cu,Liu:2011dw}, we discuss the phenomenological effect of the Pauli coupling. More precisely, we explore the role of the strength of the Pauli coupling on the form of the spectral functions.

\begin{figure}[htp]
         \centering
         \begin{subfigure}[b]{0.5\textwidth}
                 \includegraphics[width=0.8\textwidth]{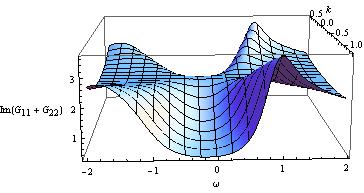}
                 \caption{$g=0$}
                 \label{fig:gull}
         \end{subfigure}%
         %~ %add desired spacing between images, e. g. ~, \quad, \qquad etc.
           %(or a blank line to force the subfigure onto a new line)
         \begin{subfigure}[b]{0.5\textwidth}
                 \includegraphics[width=0.8\textwidth]{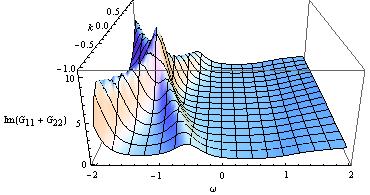}
                 \caption{$g=1$}
                 \label{fig:tiger1}
         \end{subfigure}
         ~ %add desired spacing between images, e. g. ~, \quad, \qquad etc.
           %(or a blank line to force the subfigure onto a new line)
         \begin{subfigure}[b]{0.8\textwidth}
                 \includegraphics[width=0.8\textwidth]{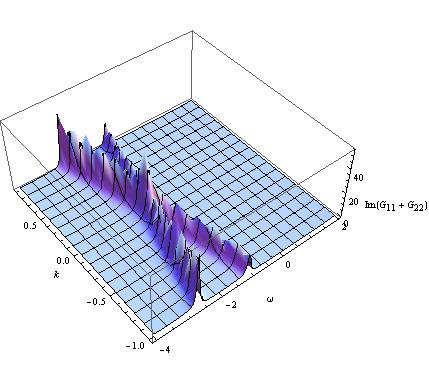}
                 \caption{$g=2$}
                 \label{fig:mouse1}
         \end{subfigure}
         \caption{Plot of the spectral function of the massless spin-3/2 field as a function of frequency $\omega$ and momentum $k$. We consider various values of the coupling $g$.}\label{fig:Pauli}
\end{figure}

Our results are summarized graphically in this subsection. First we consider, in Figure~\ref{fig:Pauli}, the role of increasing the coupling to the Pauli term on the form of the spectral function. We include three panels corresponding to $g=0,1,2$ respectively. The three-dimensional plots are particularly efficient at describing the new feature of the spectral functions. Namely, the new peak that develops and the dependence of its position on the value of $g$. Note that this Pauli induced peak is different from the ``standard'' phonino peak discussed before which is present even in the Schwarzschild-AdS case.  This feature was noted in \cite{Edalati:2010ww,Edalati:2010ge} and further interpreted as related to a shift in the spectral weight.

\begin{figure}[htp]
         \centering
         \begin{subfigure}[b]{0.45\textwidth}
                 \includegraphics[width=0.9\textwidth]{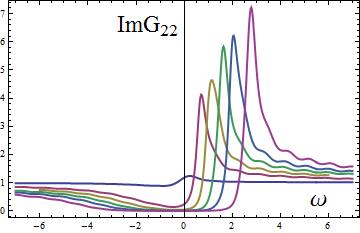}
                 \caption{This is the graph of $g=0$, $m=0$. The peaks looks ordinary. Different curves stands for different values of $k$ (from left to right: $k=0.1,1,1.5,2,2.5,3.2$)}
                 \label{fig:tiger2}
         \end{subfigure}
         ~ %add desired spacing between images, e. g. ~, \quad, \qquad etc.
           %(or a blank line to force the subfigure onto a new line)
         \begin{subfigure}[b]{0.45\textwidth}
                 \includegraphics[width=0.9\textwidth]{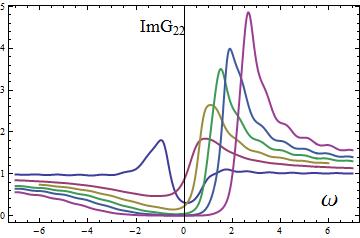}
                 \caption{This is the graph of $g=1/4$, $m=0$ (minimal coupling supergravity). The peaks began to translate. Different curves stands for different values of $k$ (from left to right: $k=0.1,1,1.5,2,2.5,3.2$).}
                 \label{fig:mouse2}
         \end{subfigure}
         \caption{Appearance of a peak for small $k$ due to nonzero $g$.}\label{Fig:New_Peak}
\end{figure}

As the coupling, $g$, increases the poles separate.  This was discovered by others in the case of spin-1/2 fields \cite{Edalati:2010ww,Edalati:2010ge,Guarrera:2011my}; here we see the same effect for the spectral functions of the spin-3/2 fields as well. In Figure \ref{Fig:New_Peak} we present a conceptual section of Figure \ref{fig:Pauli}. Namely, we consider only ${\rm Im}\,G_{22}$ for various values of the momentum (see Figure \ref{Fig:New_Peak}). As we increase $g$ to $g=1/4$ we see the appearance of the new peak.  Note that this is a very modest value of $g$ compared to $g=2$ in Figure \ref{fig:Pauli}.

\begin{figure}[htp]
         \centering
         \begin{subfigure}[b]{0.45\textwidth}
                 \includegraphics[width=0.9\textwidth]{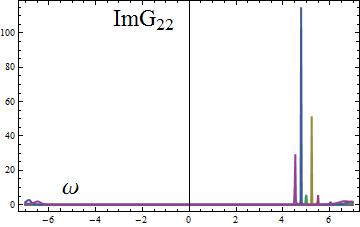}
                 \caption{Here we set $g=4.5$, $m=0$. The range of $\omega$ is $[-7,7]$.  There are peaks on the positive axis. Different curves stands for different values of $k$ (from right to left: $k=0.1,1,1.5,2,2.5,3.2$).  }
                 \label{fig:tiger3}
         \end{subfigure}
         ~ %add desired spacing between images, e. g. ~, \quad, \qquad etc.
           %(or a blank line to force the subfigure onto a new line)
         \begin{subfigure}[b]{0.45\textwidth}
                 \includegraphics[width=0.9\textwidth]{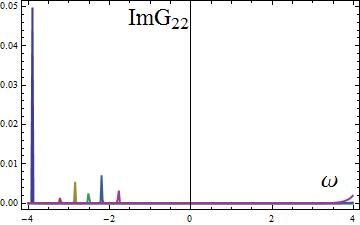}
                 \caption{Here we set $g=4.5$, $m=0$. The range of $\omega$ is shortened to be $[-4,4]$. one can see there are peaks on the negative axis. Different curves stands for different values of $k$ (from left to right: $k=0.1,1,1.5,2,2.5,3.2$).}
                 \label{fig:mouse3}
         \end{subfigure}
         \caption{Changing the scale of $\omega$ magnifies the existence of further peaks.  Note the difference in the scale (four order of magnitude).}\label{Fig:More_Peaks}
\end{figure}

In Figure \ref{Fig:More_Peaks} we consider a large value of the Pauli coupling, $g=4.5$, and show explicitly the existence of a whole hierarchy of peaks in the spectral functions. The left panel of Figure \ref{Fig:More_Peaks} shows the spectral function with easily distinguishable peaks for $4<\omega<6$; the right panel of Figure \ref{Fig:More_Peaks} is a zoom that shows the presence of peaks in the spectral function for $-4 \le \omega \le -2$. The peaks in the right panel are not visible in the zoom-out left panel. Note that the values range over four orders of magnitude.

\begin{figure}[htp]
\includegraphics[width=0.4\textwidth]{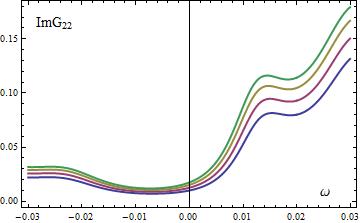}\\
\caption{Very small frequency range for the gravitino in RN-AdS${}_4$, with $m=1, g=1/4$. The values of the momenta are $k=1.1, 1.2, 1.3$ and $1.4$ from lower to higher values of the spectral functions.}\label{Fig:Gravitino_massive}
\end{figure}

We remark, again,  that in the supergravity case the large frequency behavior of the spectral function is dictated by conformal invariance and therefore is divergent. However, there is a structure for very small frequencies which we show  in Figure \ref{Fig:Gravitino_massive} and was already discussed in \cite{Gauntlett:2011mf,Gauntlett:2011wm}.

%%%%%%%%%%%%%%%%%%%%%%%%%%%%%%%%%%%%%%%%%%%%%%%%%%%
\subsubsection{Quantum critical behavior}

Via the AdS/CFT correspondence, the results of this section can be translated to a class of strongly-coupled (2+1)-dimensional field theories at zero temperature and finite $U(1)$ charge density. This is a particularly interesting situation from the condensed matter point of view \cite{sachdev2011}. In this subsection we discuss the quantum critical behavior highlighted first in \cite{Faulkner:2009wj} for the case of spin-1/2 fields. Accordingly, we consider throughout in this subsection the zero temperature limit and vanishing frequency $\omega$.

\begin{figure}[htp]
         \centering
         \begin{subfigure}[b]{0.45\textwidth}
                 \includegraphics[width=0.9\textwidth]{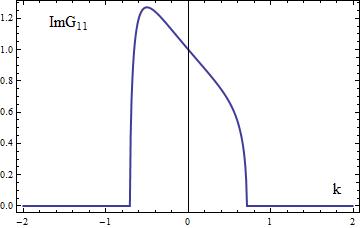}
                 \caption{Zero temperature spectral function for $m=0$ with no Pauli term $(g=0)$.}
                 \label{fig:tiger4}
         \end{subfigure}
         ~ %add desired spacing between images, e. g. ~, \quad, \qquad etc.
           %(or a blank line to force the subfigure onto a new line)
         \begin{subfigure}[b]{0.45\textwidth}
                 \includegraphics[width=0.9\textwidth]{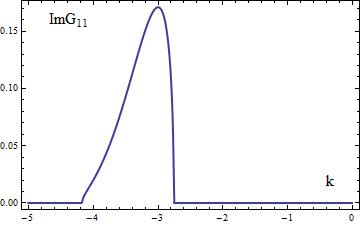}
                 \caption{Zero temperature spectral function for $m=0$ with Pauli coupling ($g=0.5$).}
                 \label{fig:mouse4}
         \end{subfigure}
         \caption{Dependence of the spectral function and its critical quantum behavior on the coupling $g$.}\label{Fig:QCB_RN}
\end{figure}

For a generic massless spin-3/2 field, one can easily see that there exists critical behavior. Allowing in this case a Pauli (dipole) coupling (for example, $g=0.5$) and keeping the mass vanishing the peak persists as we show in Figure \ref{Fig:QCB_RN}. In the case of zero temperature one anticipates potential application to quantum critical behavior on the dual field theory side. When $\omega=0$, the horizon condition is given by (\ref{eq:omega=0hc}). This condition leads to a peak  in the spectral function which we show in the left panel of Figure \ref{Fig:QCB_RN}.  In the right panel we show that the peak persists as one increases the Pauli coupling, $g$.  It is also clear in Figure \ref{Fig:QCB_RN} that a non-vanishing value of $g$ shifts the position of the peak and further affects its form.

\begin{figure}[htp]
         \centering
         \begin{subfigure}[b]{0.45\textwidth}
                 \includegraphics[width=0.9\textwidth]{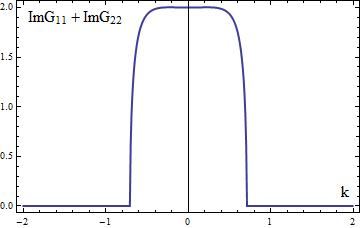}
                 \caption{Zero temperature spectral function for $m=0$ with no Pauli coupling ($g=0$).}
                 \label{fig:tiger5}
         \end{subfigure}
         ~ %add desired spacing between images, e. g. ~, \quad, \qquad etc.
           %(or a blank line to force the subfigure onto a new line)
         \begin{subfigure}[b]{0.45\textwidth}
                 \includegraphics[width=0.9\textwidth]{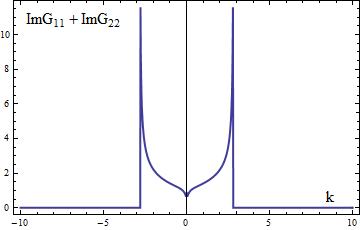}
                 \caption{Zero temperature spectral function for $m=0$ with Pauli coupling ($g=0.1$).}
                 \label{fig:tiger6}
         \end{subfigure}
         \begin{subfigure}[b]{0.45\textwidth}
                 \includegraphics[width=0.9\textwidth]{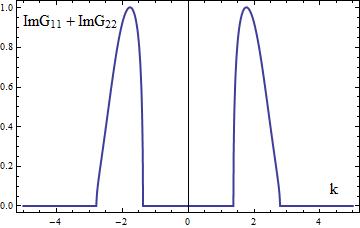}
                 \caption{Zero temperature spectral function for $m=0$ with Pauli coupling ($g=0.3$).}
                 \label{fig:tiger7}
         \end{subfigure}
         \begin{subfigure}[b]{0.45\textwidth}
                 \includegraphics[width=0.9\textwidth]{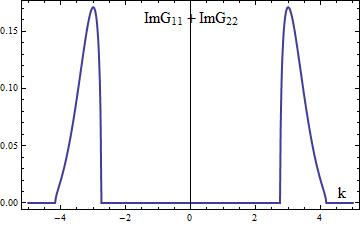}
                 \caption{Zero temperature spectral function for $m=0$ with Pauli coupling  ($g=0.5$).}
                 \label{fig:mouse5}
         \end{subfigure}
         \caption{ Shifting of peaks in spectral functions induced by increasing the Pauli coupling, $g$.}\label{Fig:Shift}
\end{figure}

Let us study analytically the dependence on $g$ of the position of the peak in the spectral function. Generally, the cutoff  momentum is given by the roots of $\kappa_1$ and $\kappa_2$ given in (\ref{Eq:Frequencies}).  For $\xi_+$ ($G_{11}$)
\begin{equation}
k_c=\frac{u^--u^+ \pm \sqrt{(u^+ +u^-)^2-4 m^2 L^2}}{2}=-2 g \sqrt{2 d (d-1)} \pm \sqrt{2 e^2 L^2 - m^2 L^2}.\label{Eq:k_G11}
\end{equation}
For $\xi_-$ ($G_{22}$)
\begin{equation}
k_c=\frac{u^+-u^- \pm \sqrt{(u^+ +u^-)^2-4 m^2 L^2}}{2}=2 g \sqrt{2 d (d-1)} \pm \sqrt{2 e^2 L^2 - m^2 L^2}.\label{Eq:k_G22}
\end{equation}
The above equations generalize a similar result in Ref.~\cite{Faulkner:2009wj} (see Eqs.~(68) and (69) there) by including the coupling of the of the Pauli term $g$. Now we can track the role that the Pauli term $\Fslash$  and its coupling, $g$, plays in the cut off  momentum; they shift the cut momentum, $k_c$, with respect to the value in a theory without Pauli interaction. Moreover, from the above equations, we notice that the cut off momentum is shifted to the left for a non-zero $g$ in $G_{11}$ and shifted to the right for $G_{22}$. This is precisely the behavior we see graphically in Figure~\ref{Fig:Shift}.

Note, however, that in (\ref{Eq:k_G11}) and (\ref{Eq:k_G22}), the special value corresponding to supergravity does not allow the occurrence of quantum critical behavior%
\footnote{It would be interesting to determine the effective value of $k_c$ in the top down models presented in \cite{Bah:2010yt,Bah:2010cu,Liu:2011dw}.}.
Namely, since in supergravity we have:
\begin{eqnarray}
\frac{u^+ + u^ -}{2 m L }=\sqrt{\frac{d-2}{d-1}}<1,
\end{eqnarray}
then, $\kappa_1$ and $\kappa_2$ are real, and there is no in-falling condition to be imposed. It is easy to check that for finite temperature there is no critical behavior either.

Let us finish our analysis by listing what we consider the most salient features of the zero-temperature spectral functions. The cut off momentum, $k_c$,  is related to the ability to set horizon conditions  and determines the potential peak in the spectral function. As we have shown algebraically, the peak exists only for  $\sqrt{2}e > m$. That is, it is a behavior strongly related to the value of the charge. Increasing the coupling of the Pauli term, $g$, induces a shift in $k_c$ and, correspondingly, in the position of the peak. In Figure~\ref{Fig:Shift}, we show that the peaks get shifted as we increase the coupling to the Pauli term. Note that ${\rm Im} \, G_{11}$ and ${\rm Im}\, G_{22}$ shift in opposite directions as the analytic prediction  (\ref{Eq:k_G11}) and (\ref{Eq:k_G22}) indicates, and finally the whole spectral function (${\rm Im}\, G_{11} + {\rm Im} \,G_{22}$) gets separated.

\section{Conclusions}

We have provided a comprehensive approach to various aspects of solving for the spin-3/2 field equations in asymptotically AdS spacetimes. We have considered the general minimally coupled plus Pauli term equation for spin-3/2 field with arbitrary $m_1$ and $m_2$ values, (\ref{Eq:RS}), as well as paid particular attention to  the supergravity case which corresponds to specific values of the couplings, (\ref{Eq:Gravitino}).

We expect  our work to serve as a spring board to applications to problems inspired by condensed matter situations and as a guide to the main results and methods in the field, some of which we have improved and clarified in this manuscript. We hope to have elucidated the conditions for emergence of certain salient features of the spectral functions. For example, we have discussed specific properties  which are present in the general case (arbitrary couplings) but absence in the particular case of supergravity. We also paid particular attention to the zero temperature case which is relevant for  quantum critical behavior on the dual field theory side.

We have established agreement and connections to various works, from the original investigations limited to spin-3/2 in AdS, to the more recent discussions of spin-3/2 in the RN-AdS background in the context of supergravity \cite{DeWolfe:2011aa,DeWolfe:2012uv}. For example, we reproduced in full detail the absence of Fermi surface for this case  \cite{Belliard:2011qq} and exhaustively discussed the effects associated with the Pauli term by expanding on previous results \cite{Edalati:2010ww,Edalati:2010ge,Guarrera:2011my}.

On a more conceptual note, we have demonstrated explicitly that the equations for general spin-3/2 fields reduce to Dirac-like equations under the appropriate decomposition and judicious choice of gauge-fixing when appropriate. We have also presented group-theoretic arguments linking various gauge choices in the literature.

We finish by highlighting some interesting open problems. Spectral functions are, almost by definition, related to the quasinormal modes. Recently two works have provided  analytic expressions for quasinormal models in the asymptotically large frequency region for spin-1/2 \cite{Arnold:2013gka} and spin-3/2 fields \cite{Arnold:2013zva}. It would be interesting to understand in more analytic terms the structure of quasinormal modes for spin-3/2 fields as well as using that information to better understand spectral functions. We took a very modest step in this direction in Appendix~\ref{App:flow}, but much more remains to be done. It is also clear that this information will play a central role in more general non-equilibrium time-dependent phenomena involving spin-3/2 fields.

Another interesting problem that remains open is the calculation of spectral functions in full consistent truncations of IIB and 11-dimensional supergravity compactified on 5-dimensional and 7-dimensional Sasaki-Einstein spaces, respectively. The complete fermionic reduction is known explicitly thanks to \cite{Bah:2010cu,Bah:2010yt,Liu:2011dw}. Finally we point out the interesting structure of the fermionic spectral function in  $p$-wave superconductors  \cite{Gubser:2010dm,Zayas:2011dw}. In view of the rich structure observed in those works, it would be interesting to extend the analysis to include the spin-3/2 field in those backgrounds.

\section*{Acknowledgments}
We thank H. Liu and J. Sonner for correspondence. L. PZ  acknowledges the hospitality at APCTP, Korea and ICTP, Italy where part of this work was done. Z.Y is thankful to the USTC/Michigan Undergraduate Research Opportunity Program that made possible this collaboration and talks with Z. Yin. This research was
supported in part by the Department of Energy under grant DE-SC0007859.

%%%%%%%%%%%%%%%%%%%%%%%%%%%%%%%%%%%%%%%%%%%%%%%%%%%%%%%%%%%%%%%%%%%%%%%%%%%%%%%%%%%%%%%%%%%%%%%%%
\appendix
\section{Conventions}\label{App:Conventions}

We denote the Dirac matrices by $\tilde{\gamma}^M$ in curved spacetimes and by $\gamma^M$ in flat spacetimes. The symbol ${\partial}_M$ refers to the ordinary partial derivative, ${\nabla}_M$ refers to the gravitational covariant derivative, and ${D}_M$ refers to the full gauge and gravitational covariant derivative.  We work in $D=d+1$ dimensions and use $M,N,\ldots=0,\ldots,d$ to denote bulk indices. $\mu,\nu,\ldots=0,\ldots,d-1$ to denote non-radial space-time indices, and $i,j=1,\ldots,d-1$ to denote non-radial space indices.  We commonly refer to $x^0$ as $t$ and $x^d$ as $r$.

We consider the general metric of Eq.~(\ref{Eq:Metric}) throughout:
\begin{equation}
ds_{d+1}^2=-e^{2A}dt^2+e^{2 B} d\vec x_{d-1}^2+e^{2 C} dr^2.
\end{equation}
The non-vanishing Christoffel symbols are
\begin{equation}
\Gamma^t{}_{tr}=A',\qquad\Gamma^i{}_{jr}=\delta^i_jB',\qquad\Gamma^r{}_{rr}=C',
\qquad
\Gamma^r{}_{tt}=-g^{rr}g_{tt}A',\qquad\Gamma^r{}_{ij}=-g^{rr}g_{ij}B',
\end{equation}
where primes denote derivatives with respect to $r$.  The non-vanishing Ricci components are
\begin{eqnarray}
R_t^t&=&-g^{rr}(A''+A'(A'+(d-1)B'-C')),\nn\\
R_x^x&=&-g^{rr}(B''+B'(A'+(d-1)B'-C')),\nn\\
R_r^r&=&-g^{rr}(A''+A'(A'-C')+(d-1)(B''+B'(B'-C'))),
\end{eqnarray}
and $R=R_t^t+(d-1)R_x^x+R_r^r$.

For this metric, the curved and flat Dirac matrices are related by
\begin{equation}
\tilde\gamma^t=e^{-A}\gamma^t, \quad  \tilde\gamma^i=e^{-B}\gamma^i, \quad \tilde\gamma^r= e^{-C}\gamma^r.
\end{equation}
The spin connection matrix $\omega_M{}^{AB}\tilde\gamma_{AB}$ is as follows:
\begin{eqnarray}
\omega_t=2A' \tilde\gamma_{t} \tilde\gamma^r, \qquad
\omega_i=2B'\tilde\gamma_i \tilde\gamma^r.
\end{eqnarray}
The Dirac operator acting on a spin-1/2 spinor is given by
\begin{equation}
\tilde\gamma^M D_M=\tilde\gamma^\mu(\partial_\mu-ieA_\mu)+\tilde\gamma^r\left(
\partial_r-ieA_r+\frac{1}{2}(A'+(d-1)B')\right).
\end{equation}
Given the form of the metric, and the fact that it depends only on the radial coordinate, it is possible to simplify the equations by effectively eliminating the role of the spin connection. Namely, defining $\psi=e^{-\frac{A+(d-1)B}{2}}\hat\psi$, leads to an equation for the rescaled field $\hat\psi$ that removes the spin connections terms.  Thus
\begin{equation}
\tilde\gamma^M D_M\hat\psi=\tilde\gamma^M(\partial_M-ieA_M)\hat\psi.
\end{equation}

Although we work in $D$-dimensions, we maintain isotropy of the $d-1$ space dimensions.  Hence we will not need an explicit basis of $D$-dimensional Dirac matrices.  Instead, it is sufficient to take
\begin{eqnarray}
\gamma^0&=&\left(
             \begin{array}{cccc}
               0 & 0 & 0 & 1 \\
               0 & 0 & -1 & 0 \\
               0 & 1 & 0 & 0 \\
               -1 & 0 & 0 & 0 \\
             \end{array}
           \right), \quad
\gamma^1=\left(
             \begin{array}{cccc}
               0 & 0 & 0 & 1 \\
               0 & 0 & 1 & 0 \\
               0 & 1 & 0 & 0 \\
               1 & 0 & 0 & 0 \\
             \end{array}
           \right) ,  \quad
\gamma^r=\left(
             \begin{array}{cccc}
               1 & 0 & 0 & 0 \\
               0 & 1 & 0 & 0 \\
               0 & 0 & -1 & 0 \\
               0 & 0 & 0 & -1 \\
             \end{array}
           \right). \nonumber
\end{eqnarray}

%%%%%%%%%%%%%%%%%%%%%%%%%%%%%%%%%%%%%%%%%%%%%%%%%%%%%%%%%%%%%%%%%%%%%%%%%%%%%%%%%%%%%%%%%%%%%%%%%%%%%%%%%%%%%55
\section{A flow equation for the Green's function}
\label{App:flow}

We review the  approach we followed in solving the equations of motion for spin-3/2 fields in this appendix. We elaborate upon and generalize an observation made in \cite{Iqbal:2008by,Liu:2009dm} about the Green's function calculation. Namely, the fact that, in some cases, an equation for the evolution of the Green's function can be written explicitly. Effectively, we transform the second order equation of motion into a non-linear first order equation. The advantage of this transformation is that it provides a numeric improvement because the horizon behavior is no longer oscillatory. In particular, oscillatory behavior such as $\psi=(1-r)^{-{i \omega}/{\nu }}$  is transformed into $\xi=i$, which is simply constant at the horizon. This change in the boundary conditions results in a substantial computational advantage.  Essentially we turn the problem of determining two coefficients $A$ and $B$ into an initial value  problem with the subsequent  evolution of a function.  This method turns out to be quite general and it is based  on a property of a class of nonlinear differential equations (Riccati equations) which can be also presented as second order differential equations.

Namely, following  \cite{bender1999advanced}, given a Riccati-type equation
\begin{eqnarray}
y'=\alpha y^2+\beta y + \gamma,
\end{eqnarray}
we can introduce a new function $\omega$  as
\begin{equation}
y= Q \frac{\omega'}{\omega}.
\end{equation}
Then by choosing $Q=-1/\alpha$, the original Riccati equation becomes a second order homogeneous differential equation:
\begin{eqnarray}
\omega''=(\beta-(\log Q)')\omega'+\frac{\gamma \omega}{Q}.
\end{eqnarray}

%%%%%%%%%%%%%%%%%%%%%%%%%%%%%%%%%%%%%%%%%%%%%%%%%%%%%%%%%%%%%%%
\subsection{Scalar Field}
%%%%%%%%%%%%%%%%%%%%%%%%%%%%%%%%%%%%%%%%%%%%%%%%%%%%%%%%%%%%%%%
To display the universality of the method we employ throughout the paper we first describe it in the case of the scalar Klein-Gordon equation. Consider a metric of the form
\begin{equation}
ds^2=-e^{2A}dt^2+e^{2 B} d\vec x^2+e^{2 C} dr^2.
\end{equation}
The Klein-Gordon equation for a massive scalar is
\begin{equation}
\phi''+(A'+(d-1)B'-C')\phi' + (e^{2C-2B}(u^2-\vec k^2)-e^{2 C}m^2) \phi=0,
\label{eq;KGe}
\end{equation}
where
\begin{equation}
u=e^{B-A}(\omega + e A_t).
\end{equation}
Defining
\begin{equation}
\xi=\frac{e^{B-C}}{k+u}\frac{\phi'}{\phi},
\end{equation}
the field equation becomes
\begin{equation}
\xi'=-(k+u)e^{C-B}\xi^2-(A'+(d-2)B'+\partial_r\log(k+u))\xi-e^{C-B}(u-k)+\frac{e^{B+C}m^2}{k+u}.
\label{eqn:KGe}
\end{equation}
This is the flow equation we are searching for.

The boundary conditions at the horizon determine the type of Green's function we are computing. First, let us translate the asymptotic behavior of $\phi$ into the asymptotic behavior of $\xi$:
\begin{eqnarray}
\xi&=&\frac{e^{B-C}}{k+u}\frac{\phi'}{\phi}\rightsquigarrow \frac{1}{k+\bar\omega}\frac{B \Delta_+ r^{\Delta_+-1}+ A \Delta_- r^{\Delta_--1}}{B r^{\Delta_+}+A r^{\Delta_-}}\rightsquigarrow\frac{\Delta_+-\Delta_-}{k+\bar\omega} G_R r^{\Delta_+-\Delta_--1}+\frac{\Delta_-}{k+\bar\omega}\frac{1}{r}, \nonumber \\
G_R&=&\lim _{ \epsilon \rightarrow 0 }{\frac{k+\bar\omega}{\Delta_+-\Delta_-} \epsilon^{-\Delta_++\Delta_-+1}\xi\Big|_{r=\epsilon}},
\end{eqnarray}
where $\Delta\pm=\frac{d}{2}\pm \sqrt{(\frac{d}{2})^2+(m L)^2}$, and we have extracted the finite terms in the limit (which is real and thus does not influence the spectral function).

For the horizon condition, we need to select the in-falling wave solution of $\phi$. For a finite temperature horizon, the field is asymptotically:
\begin{equation}
\partial_z^2\phi+\frac{1}{z}\partial_z\phi + \frac{1}{\nu z}(u^2- k^2-(m L)^2)\phi=0,
\end{equation}
where $z, \nu$ are defined in section 4.
When $\omega$ is not zero, the equation and solution are:
\begin{eqnarray}
\partial_z^2\phi+\frac{1}{z}\partial_z\phi + \frac{\omega^2}{\nu^2 z^2}\phi=0, \quad \Longrightarrow\quad
\phi=z^{\pm {i \omega}/{\nu }}.
\end{eqnarray}
The  in-falling wave solution is $\phi=z^{-{i \omega}/{\nu }}=(1-r)^{-{i \omega}/{\nu }}$. Hence, $\xi$ becomes
\begin{equation}
\frac{e^{B-C}}{k+u}\frac{\phi'}{\phi}=i.
\end{equation}
At finite temperature  and for vanishing $\omega$, the equation is:
\begin{eqnarray}
\partial_z^2\phi+\frac{1}{z}\partial_z\phi =0, \quad \Longrightarrow\quad
\phi=C_1+C_2 \log(z).
\end{eqnarray}
This solution does not behave like a wave. Hence there is no sensible horizon condition.
Extra care is needed when the temperature is zero. First we consider  the case $\omega\neq 0$. The field equation near the horizon now becomes:
\begin{eqnarray}
\partial_z^2 \phi + \frac{2}{z}\partial_z\phi+\frac{\tilde \omega^2}{z^4} \phi=0,
\end{eqnarray}
where
\begin{equation}
\tilde \omega = \frac{\omega}{d(d-1)}, \qquad z=1-r.
\end{equation}
Using  these definitions the equation can be rewritten as
\begin{eqnarray}
\tphi''+\frac{\tilde \omega^2}{z^4}\tphi&=&0, \qquad {\rm with} \quad \tphi=z \phi.
\end{eqnarray}
Since we are considering the behavior around $z=0$ and the term ${\tilde \omega^2}/{z^4}$ has no turning  points in this regime, we can use the WKB approximation around $z\approx 0$. Using WKB, we get the solution of $\tphi$ to be $C z e^{\pm {i \tilde\omega}/{z}}$.  The in-falling wave condition now becomes
\begin{eqnarray}
\phi\rightarrow C e^{ {i \tilde\omega}/{z}}, \qquad \xi=\frac{e^{B-C}}{k+u}\frac{\phi'}{\phi}\rightarrow i.
\end{eqnarray}
This case of zero temperature is particularly relevant for condensed matter applications, as it constitutes a window into quantum transitions. Now, consider the $\omega=0$ case. In this case the  field equation  becomes
\begin{eqnarray}
\partial_z^2\phi + \frac{2}{z}\partial_z \phi + \frac{U^2}{z^2}\phi=0, \quad
U^2= \frac{u^2-\vec k^2-(m L)^2}{d (d-1)}, \quad u= e L \sqrt{2}.
\end{eqnarray}
The solution to this equation is $\phi= z^n$, and $n^2+n+U^2=0$.  To get oscillatory behavior, we need a complex root, which means $U^2 \leq {1}/{4}$, and the solution is $\phi=z^{({-1 \pm i \sqrt{4 U^2-1}})/{2}}$.  The in-falling wave corresponds to  $z^{({-1 - i \sqrt{4 U^2-1}})/{2}}$, leading to
\begin{equation}
\xi=\frac{e^{B-C}}{k+u}\frac{\phi'}{\phi}\rightarrow \frac{\sqrt{d(d-1)}\left(\frac{1}{2}+ i \sqrt{U^2-\frac{1}{4}}\right)}{k+u}.
\end{equation}
When $U^2 \geq {1}/{4}\Leftrightarrow 2e^2L^2 \geq \frac{d(d-1)}{4} +k^2+(m L)^2$, we have $\mbox{Im\,}G_R \neq 0$; when $2e^2L^2 \leq \frac{d(d-1)}{4} +k^2+(m L)^2$, we find instead $\mbox{Im\,}G_R=0$. This quantum critical behavior was also pointed out in Ref.~\cite{Faulkner:2009wj} using a different method.

%%%%%%%%%%%%%%%%%%%%%%%%%%%%%%%%%%%%%%%%%%%%%%%%%%%%%%%%%%%%%%%%%
\subsection{Green's function for a spin-1/2 (spin-3/2)  field using the flow equation}

As we can transform the field equation of motion into a flow equation, we can also transform our flow equation back to a field equation. This give us another aspect to the analysis our flow equation.

The flow equation is
\begin{equation}
\partial_r \xi=e^{C-B} (k_1- u^-)+ 2 e^C m \xi - e^{C-B} (k_1+ u^+)\xi^2.
\end{equation}
Set $\xi = \frac{e^{B-C}}{k+u^+} \frac{\psi'}{\psi}$, we obtain a second order equation for $\psi$ as:
\begin{equation}
\psi''+(-2 e^C m + B'- C' -\partial_r \log(k+u^+))\psi'- e^{2C-2B}(k+ u ^+)(k-u^-)\psi=0.
\end{equation}
Note that in this case the mass term is different from the scalar case, (\ref{eqn:KGe}). In that case the mass shows up as $m^2$ and is coupled with $\phi$, while in this case, the mass shows up as $m$ and is coupled with $\psi'$. Later on we will see that it is this difference that leads to the crucial role  the mass plays in spinor fields as opposed to in scalar fields. Let us, again, consider the horizon condition and boundary behavior. Around a finite temperature horizon the solution of $\psi$ is
\begin{equation}
(1-r)^{\pm i \omega/\nu}.
\end{equation}
 This condition introduces a numerical error with respect to the condition in terms of the function $\xi$, whose horizon condition is simply $\xi=i$. For zero temperature and around the horizon, when $\omega=0$, the field equation and solution are:
\begin{eqnarray}
&&\partial_z^2\psi+\frac{1+\frac{2 m L}{\sqrt{d(d-1)}}}{z}\partial_z\psi-\frac{(k+u^+)(k-u^-)}{d(d-1)z^2}\psi=0,\\
&&\psi=z^{\frac{-m L \pm \sqrt{m^2 L^2 +(k+u^+)(k-u^-)}}{\sqrt{d(d-1)}}}.
\end{eqnarray}
Let us compare with the horizon condition in the $\xi$ variable
\begin{equation}
\xi=\begin{cases}
     \frac{m L+ \sqrt{m^2 L^2 +(k+u^+)(k-u^-)}}{k+u^+}, &m^2 L^2 +(k+u^+)(k-u^-)<0; \\
    0, & m^2 L^2 +(k+u^+)(k-u^-)\geq0.
  \end{cases}
\end{equation}
In terms of the field $\psi$, the horizon condition is
\begin{equation}
\psi=\begin{cases}
    z^{\frac{-m L - \sqrt{m^2 L^2 +(k+u^+)(k-u^-)}}{\sqrt{d(d-1)}}} , & m^2 L^2 +(k+u^+)(k-u^-)<0;\\
    0, & m^2 L^2 +(k+u^+)(k-u^-)\geq0.
  \end{cases}
\end{equation}
For finite temperature and $\omega\neq0$, we get the solution:
\begin{eqnarray}
\psi=z^{1-\frac{m L}{d(d-1)}}e^{\pm i\frac{\omega}{d(d-1)z}},
\end{eqnarray}
and the horizon condition turns to be:
\begin{equation}
\psi=z^{1-\frac{m L}{d(d-1)}}e^{+ i\frac{\omega}{d(d-1)z}}.
\end{equation}
Note that the horizon conditions all correspond to in-falling wave condition. Near the boundary, the solution of $\psi$ is
\begin{equation}
\psi(r)=r^n(A Y_n (\lambda r)+ B J_n (\lambda r)),
\end{equation}
where $\lambda^2= \bar\omega^2-\vec k ^2, \,\, n= mL +\frac{1}{2}$.  So,
\begin{equation}
G_R=\lim _{ \epsilon \rightarrow 0 }{\epsilon ^{-2 m L}\xi\Big|_{r=\epsilon} }=\lim _{ \epsilon \rightarrow 0 }{\epsilon ^{-2 m L}\frac{1}{k+\bar\omega}\frac{B J_{n-1}}{A Y_n} }\Big|_{r=\epsilon},
\end{equation}
where we have subtracted the divergent term:
\begin{equation}
r^{-2 m L}\frac{1}{k+\tilde\omega}\frac{Y_{n-1}}{Y_{n}}\Big|_{r\rightarrow 0}.
\end{equation}
This formula for the Green's function is actually very useful, and we turn to  a detailed  discussion of some of its properties.  Let us review how it reproduces previous results.  We consider the massless case in Schwarzschild-AdS. Then our equation takes the form:
\begin{eqnarray}
\psi''+\beta \psi' +\gamma \psi=0, \qquad
\nonumber \beta=\frac{f'}{f}-\frac{f'}{2\sqrt{f}(\omega+\sqrt{f}k)},  \qquad \gamma=\frac{\omega^2-k^2 f}{f^2},  \qquad f=1-r^4.
\end{eqnarray}
After replacing $\psi$ by $e^{-\int\frac{\beta}{2}}\tpsi$, this goes to
\begin{eqnarray}
\tpsi''=Q\tpsi,\quad Q=\frac{\beta'}{2}+\frac{\beta^2}{4}-\gamma.
\end{eqnarray}
For the case of $\omega$ large enough, $Q$ is negative in the range of $[0,1]$, so there are no turning points.  Hence the WKB method is valid to get an approximate solution
\begin{equation}
\tpsi\sim C \frac{e^{i\int^r_0 \sqrt{-Q(t)}d t}}{Q^{1/4}},
\end{equation}
Then the Green's function is
\begin{equation}
\lim _{ \epsilon \rightarrow 0 }{\frac{1}{k+\omega}\frac{\psi'}{\psi} }\Big|_{r=\epsilon}.
\end{equation}
Taking the imaginary part leads to\cite{Liu:2009dm}
\begin{equation}
\frac{\sqrt{-Q}}{k+\omega}\Big|_{r\rightarrow0}=\sqrt{\frac{\omega-k}{\omega+k}}.
\end{equation}
Let us now turn to the role of mass.  To simplify the discussion, we  consider pure AdS. Then our equation is:
\begin{equation}
\psi''-\frac{2 m L}{r} \psi' +(\omega^2-k^2) \psi=0.
\end{equation}
This is similar to the equation for a massless scalar field except that here the effective dimension is $2 m L + 1$. We can calculate the Green's Function as follows:
\begin{enumerate}
\item  Rescale $r$; i.e.\ let $\varsigma=r \sqrt{\omega^2-k^2}$.
\item Solve the equation at the boundary:
\begin{equation}
\psi\approx A + B \varsigma^{2 m L + 1}=A + B (\omega^2 - k^2)^{m L+1/2} r ^{2 m L+1}.
\end{equation}
\item  Get the Green's Function:
\begin{equation}
G\propto\frac{1}{k+\omega}\frac{B (\omega^2 - k^2)^{m L+1/2}}{A} \propto (\omega-k)^{m L+1/2} (\omega + k)^{m L -1/2}.
\end{equation}
\end{enumerate}
Hence we know the behavior of Green's Function is roughly $\omega^{2 m L}$ for large $\omega$.  Now we know the role of mass; it leads to a divergent behavior in the Green's Function. This effect can be seen in our numerical treatment (see Figure~\ref{Fig:Sch_Gra}).

%\subsection{Discussion about the flow equation and Classification of the Phenomenons}

We can now see that the Green's functions are purely determined by the horizon conditions around the black hole and the equations themselves regardless of the type of our particle (bosonic or fermionic).  We can regard the horizon condition as the initial value of our Green's function. Hence our results show that when the temperature is not zero, the initial values at the horizon are always $i$. For zero temperature, they share the same behavior in momentum (a cut-off).

Largely inspired by an analytic approach presented in  \cite{Gulotta:2010cu}, we now proceed to analyze the flow equation for the Green's function with the hope of clarifying some properties discussed in the main text. Since the information of the Green's function is contained in the flow equation, we need to look at this equation more carefully.

We have already mentioned the relation between $G_{11}$ and $G_{22}$. Therefore we focus on one of the equations:
\begin{equation}
e^{B-C}\partial_r \xi_-=- (k_1+ u^-)+ 2 e^B m \xi_- + (k_1- u^+)\xi_-^2.
\end{equation}
To simplify the discussion, we limit our analysis to the situation with no gauge field.  Since our focus is on the spectral function, that is, the imaginary part of the Green's function, we find it convenient to  separate  $\xi_-$ into $u+i v$. Then the flow equation can be rewritten as:
\begin{eqnarray}
 u'&=&-\left(\frac{k}{\sqrt{1-r^d}}+\frac{\omega}{1-r^d}\right)+ \frac{2 m }{r \sqrt{1-r^d}} u +\left(\frac{k}{\sqrt{1-r^d}}-\frac{\omega}{1-r^d}\right)(u^2-v^2),\nn\\
v'&=&\frac{2 m}{r \sqrt{1-r^d}}v + 2 \left(\frac{k}{\sqrt{1-r^d}}-\frac{\omega}{1-r^d}\right)u v,\nn\\
u\mid_{r\mapsto1}&=&0,\nn\\
v\mid_{r\mapsto1}&=&1.
\end{eqnarray}
We can formally solve for $v$ in terms of $u$
\begin{equation}
v=r^{2 m} \left(1+\sqrt{1-r^d}\right)^{-\frac{4 m}{d}} \exp\left(2 \int^{r}_1 \left(\frac{k}{\sqrt{1-t^d}}-\frac{\omega}{1-t^d}\right)u\, dt\right).
\end{equation}
Hence, the imaginary part of Green's function is:
\begin{equation}
{\rm Im} G_{22}=2^{-\frac{4 m}{d}} \exp\left(2 \int^{0}_1 \left(\frac{k}{\sqrt{1-t^d}}-\frac{\omega}{1-t^d}\right)u\, dt\right).
\end{equation}
Note that we have converted a complicated calculation into an integral including $u$.
There are some conclusions we can draw under mild assumptions on $u$:
\begin{enumerate}
  \item ${\rm Im} \,G_{22}$ is positive, as expected by unitarity.
  \item $k\mapsto -k, \omega\mapsto-\omega\quad \Rightarrow\quad u\mapsto-u, v\mapsto v$; hence ${\rm Im}\, G_{22} $ does not change.
  \item In the case of $m=0$, when $k=0$ or $|\omega| \gg |k|$, we can find a solution easily: $u\equiv0$, $v\equiv1$, hence ${\rm Im}\, G_{22}\equiv1$. We verified this $m=0$ behavior in, e.g., Figures \ref{Fig:Sch_RS}, \ref{Fig:RN_Gra} and \ref{Fig:New_Peak}.
\end{enumerate}

\begin{figure}[htp]
         \centering
         \begin{subfigure}[b]{0.45\textwidth}
                 \includegraphics[width=0.9\textwidth]{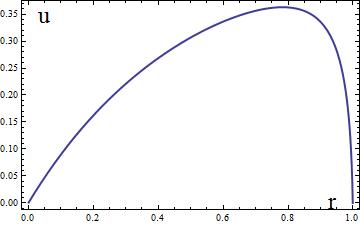}
                 \caption{Behavior of $u$ for  ($4 m \omega \gg k d $)}
                 \label{fig:tiger8}
         \end{subfigure}
         ~ %add desired spacing between images, e. g. ~, \quad, \qquad etc.
           %(or a blank line to force the subfigure onto a new line)
         \begin{subfigure}[b]{0.45\textwidth}
                 \includegraphics[width=0.9\textwidth]{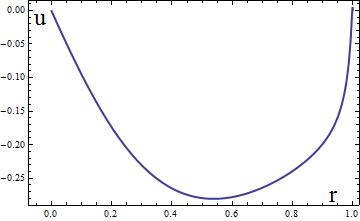}
                 \caption{Behavior of $u$ for  ($4 m \omega \ll k d $)}
                 \label{fig:mouse6}
         \end{subfigure}
         \caption{Behavior of the function $u$ in two regimes of frequencies.}\label{Fig:SF_Analytic}
\end{figure}

Let us investigate the behavior of  $u$ and $v$ near the horizon. We can expand around $r=1$, $v\rightarrow 1+ A \sqrt{\epsilon}$ and $u\rightarrow B \sqrt{\epsilon}$ (where we set $r=1-\epsilon$). The relations between $B$ and $A$ following from the flow equation are:
\begin{eqnarray}
B=\frac{4 k}{\sqrt{d}}-\frac{4 \omega}{d} A, \quad
A=-\frac{4 m}{\sqrt{d}}+\frac{4 \omega}{d}B,
\end{eqnarray}
resulting in
\begin{equation}
A=\frac{\frac{4}{\sqrt{d}}(\frac{4 k \omega}{d}- m)}{1+16 \omega^2/d^2}, \qquad B=\frac{\frac{4}{\sqrt{d}}(\frac{4 m \omega}{d}+k)}{1+16 \omega^2/d^2}.
\end{equation}
Near the  boundary $r=0$, we get that the main contribution  to $u$ is $\frac{k+\omega}{2 m -1} r + {\rm Re} \,G_{22} r^{2 m}$ and  to $v$ is ${\rm Im} G_{22}\,r^{2 m}$. Hence when $4 m \omega \gg(\ll)\, k d $, $u$ is greater (smaller) than $0$ around the horizon, and we can assume that this is the leading behavior of $u$ (see Figure~\ref{Fig:SF_Analytic}).

Based on this behavior, we can make some qualitative analysis of our Green's function. We can separate the integral into two parts: $0$ to $1-\delta$ and $1-\delta$ to $1$. So the integral turns out to be approximately: $-(2(k-\omega)\bar u + (k \sqrt{\delta} - {2 \omega}/{\sqrt{d}}) B \sqrt{{\delta}/{d}})$, where $\bar u$ is a specific value of $u$.
Therefore, our Green's function is:
\begin{equation}
{\rm Im} G_{22}=2^{-\frac{4 m}{d}} \exp\left(2(\omega-k)\bar u + \left(\frac{2 \omega}{\sqrt{d}}-k \sqrt{\delta} \right) \sqrt{\frac{\delta}{d}}B\right).
\end{equation}

\begin{figure}[htp]
\includegraphics[width=0.5\textwidth]{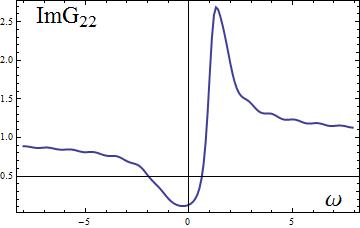}\\
\caption{Confirmation of some analytical results. For $m=0$, $k=1$, the peak in the spectral function is around $\omega=1$, and the spectral function asymptotes to  $1$.}\label{Fig:Peak_omega}
\end{figure}

Now we set $m=0$ and $k=1$, and our assumption dictates  that both $\bar u$ and $B$ are positive. So when $\omega$ is a lot  smaller than $k$, our $\mbox{Im\,}G_{22}$ is smaller than $1$. When $\omega$ is greater than $k$, our $\mbox{Im\,}G_{22}$ is greater than $1$. When $|\omega|$ is a lot greater than $k$, our previous analysis tells us that $\mbox{Im\,}G_{22}$ is $1$. It follows that the corresponding  graph will have a peak around $k$, and also a hole. This is the exact result we got previously. Also note that in Figure \ref{Fig:Peak_omega} the peak corresponds approximately to $\omega \sim 1$ and that the spectral function goes to unity for large absolute value of $\omega$.
\noindent
\bibliographystyle{JHEP}
\bibliography{AdS_CFT}
\end{document}